\documentclass[letterpaper,11pt]{article}

\usepackage{cite}
\usepackage{latexsym}
\usepackage{amsmath, amsthm, amssymb}
\usepackage{mathrsfs}
\usepackage[dvips,dvipdf]{graphicx}
\usepackage{subfigure}
\usepackage{algorithm}
\usepackage{algorithmic}
\usepackage{multirow}
\usepackage{bbm}

\usepackage{pict2e,picture} 
\usepackage{tikz}   
\usepackage{pifont} 

\newcommand{\A}{\mathbf{A}}
\renewcommand{\a}{\mathbf{a}}

\newcommand{\Bt}{\widetilde{\mathbf{B}}}
\renewcommand{\b}{\mathbf{b}}
\newcommand{\bt}{\widetilde{\mathbf{b}}}
\newcommand{\C}{\mathbf{C}}
\newcommand{\D}{\mathbf{D}}
\newcommand{\Dh}{\widehat{\mathbf{D}}}
\renewcommand{\d}{\mathbf{d}}

\newcommand{\F}{\mathbf{F}}
\newcommand{\G}{\mathbf{G}}
\newcommand{\I}{\mathbf{I}}
\newcommand{\M}{\mathbf{M}}

\newcommand{\Mt}{\widetilde{\mathbf{M}}}
\newcommand{\N}{\mathbf{N}}

\newcommand{\Q}{\mathbf{Q}}

\newcommand{\R}{\mathbf{R}}

\renewcommand{\S}{\mathbf{S}}
\newcommand{\s}{\mathbf{s}}

\renewcommand{\u}{\mathbf{u}}
\renewcommand{\v}{\mathbf{v}}

\newcommand{\w}{\mathbf{w}}
\newcommand{\wt}{\widetilde{\mathbf{w}}}
\newcommand{\wts}{\widetilde{w}}
\newcommand{\X}{\mathbf{X}}
\newcommand{\x}{\mathbf{x}}

\newcommand{\xh}{\widehat{\mathbf{x}}}

\newcommand{\y}{\mathbf{y}}
\newcommand{\z}{\mathbf{z}}
\newcommand{\yt}{\widetilde{\mathbf{y}}}

\newcommand{\0}{\mathbf{0}}
\newcommand{\1}{\mathbf{1}}
\newcommand{\Hc}{\mathbf{H}}
\newcommand{\tHc}{\widetilde{\mathbf{H}}}

\newcommand{\mLambda}{\mathbf{\Lambda}}
\newcommand{\mOmega}{\mathbf{\Omega}}
\newcommand{\mOmegab}{\overline{\mathbf{\Omega}}}

\newcommand{\Out}[1]{\mathcal{O}\left( #1 \right)}
\newcommand{\In}[1]{\mathcal{I}\left( #1 \right)}

\newcommand{\src}[1]{\mathrm{Src}(#1)}
\newcommand{\dst}[1]{\mathrm{Dst}(#1)}
\newcommand{\rx}[1]{\mathrm{Rx}(#1)}
\newcommand{\tx}[1]{\mathrm{Tx}(#1)}
\newcommand{\Dy}{\Delta \y}
\newcommand{\Dyt}{\Delta \widetilde{\y}}
\newcommand*\circled[1]{\tikz[baseline=(char.base)]{
            \node[shape=circle,draw,inner sep=2pt] (char) {#1};}}

\newtheorem{thm}{Theorem}[section]

\newtheorem{lem}[thm]{Lemma}
\newtheorem{prop}[thm]{Proposition}

\newtheorem{rem}{Remark}

\newcommand{\mtrx}[1]{\mbox{$\left[\begin{array} #1 \end{array}\right]$}}

\newcommand{\diag}[1]{\mathrm{Diag}\left\{ #1 \right\}}
\newcommand{\diagv}[1]{\mathrm{diag}\left\{ #1 \right\}}
\newcommand{\etal}{{\em et al. }}
\newcommand{\ind}{\mathbbm{1}}

\begin{document}

\title{A Distributed Newton Approach for Joint Multi-Hop Routing and Flow Control: Theory and Algorithm}


\author{Jia Liu$^{\dag}$ \mbox{\hspace{0.4cm}} Hanif D. Sherali$^{\ddag}$
\\ $^{\dag}$Department of Electrical and Computer Engineering, The Ohio State University
\\ $^{\ddag}$Grado Department of Industrial and Systems Engineering, Virginia Tech
}

\date{}
\maketitle

\begin{abstract}
The fast growing scale and heterogeneity of current communication networks necessitate the design of distributed cross-layer optimization algorithms.
So far, the standard approach of distributed cross-layer design is based on dual decomposition and the subgradient algorithm, which is a first-order method that has a slow convergence rate.
In this paper, we focus on solving a joint multi-path routing and flow control (MRFC) problem by designing a new distributed Newton's method, which is a second-order method and enjoys a quadratic rate of convergence.
The major challenges in developing a distributed Newton's method lie in decentralizing the computation of the Hessian matrix and its inverse for both the primal Newton direction and dual variable updates.
By appropriately reformulating, rearranging, and exploiting the special problem structures, we show that it is possible to decompose such computations into source nodes and links in the network, thus eliminating the need for global information.
Furthermore, we derive closed-form expressions for both the primal Newton direction and dual variable updates, thus significantly reducing the computational complexity.
The most attractive feature of our proposed distributed Newton's method is that it requires almost the same scale of information exchange as in first-order methods, while achieving a quadratic rate of convergence as in centralized Newton methods.
We provide extensive numerical results to demonstrate the efficacy of our proposed algorithm.
Our work contributes to the advanced paradigm shift in cross-layer network design that is evolving from first-order to second-order methods.
\end{abstract}


\section{Introduction}
The scale of current communication networks has been growing rapidly in recent years as the demand for data access continues to increase exponentially.
As a result, maintaining a centralized network control unit has become increasingly difficult or even undesirable in many situations, such as in multi-path routing and congestion control in the Internet, sensor networks, or ad hoc networks.
In such cases, distributed algorithms are not only desirable, but also necessary.

In the literature, the standard distributed approach for jointly optimizing multi-path routing and flow control (MRFC) is based on the Lagrangian dual decomposition framework and the subgradient method for dual updates, where subgradients (based on first-order supporting of the dual function) are used as search directions (see, e.g., \cite{Chiang06:Decompose,Kelly98:NUM,Low04:Distr_Algs} and references therein).
The dual decomposition framework and the subgradient approach are also related to the celebrated throughput-optimal ``back-pressure'' algorithm, which is the foundation of a large number of interesting routing and scheduling schemes (see, e.g., \cite{Tassiulas92:BackPressure,Neely05:BackPressure} and many other later works).\footnote{It can be shown by some appropriate scaling that the queue lengths can be interpreted as the dual variables in the dual decomposition framework -- see Section~\ref{sec:Subgradient} for a more detailed discussion.}
However, despite its simplicity and theoretical appeal, the subgradient method does not work well in practice.
This is because the subgradient method, a first-order approach in nature, has a slow rate of convergence (and typically exhibits a zigzagging phenomenon if the objective function is ill-conditioned) and is very sensitive to step-size selections.
These limitations motivate us to design a distributed Newton algorithm for the MRFC problem.
The fundamental rationale of using a distributed Newton method is that, being a second-order approach, a distributed Newton method would exploit both first and second-order information (more precisely, the gradient and Hessian of the underlying problem) in determining search directions.
As a result, a properly designed distributed Newton algorithm also enjoys the same {\em quadratic rate of convergence} as in the classical Newton type methods \cite{Bazaraa_Sherali_Shetty_93:NLP,Boyd_Vandenberghe04:Cnvx_Opt}.

However, due to a number of major technical challenges, research on second-order based distributed algorithms for network optimization is still in its infancy and results are rather limited.
To our knowledge, only a handful of works exist in this area (see Section~\ref{sec:Related} for more detailed discussions).
The first major technical challenge is that the computation of primal Newton direction in a second-order method typically requires taking the inverse of the Hessian matrix of the underlying problem (or solving a linear equation system), which is not always easy by itself for large-scale problems, let alone being done in a distributed fashion.
Therefore, when designing distributed second-order algorithms, one needs to figure out how to decompose the inverse of Hessian matrix and distribute each piece to
each network entity (i.e., a node or a link) in such a way that each piece can be computed using only local information or via a limited scale of information exchange between network entities.
This task is not trivial except in some special problems, and certainly not in our case.
The second major challenge is that, as we shall see later, the computation of dual variables (which represent certain pricing information) in a distributed second-order method also requires taking the inverse of some complex transformation of the Hessian matrix and
needs global information.
In fact, how to compute the dual variables in a distributed way has remained largely unaddressed until very recently, when some interesting ideas based on Gaussian belief propagation (to avoid direct matrix inversion \cite{Bickson09:GaBP}) or matrix splitting (to iteratively compute the matrix inverse \cite{Woznicki01:Mtrx_Split}) were proposed for some relatively simpler network optimization problems \cite{Bickson08:DNewton,Bickson09:DNewton,Jadbabaie09:DNewton,Wei10:DNewton}.
However, it remains unclear whether these ideas can be readily extended to more complex cross-layer optimization problems.
Therefore, our goals in this work are centered around tackling these difficulties.
The main results and contributions in our work are as follows:
\begin{itemize}
\item We show that, by appropriately reformulating and rearranging, it is possible to expose a block diagonal structure in the Hessian matrix of the MRFC problem.
    As a result, the Hessian matrix can be decomposed with respect to source nodes and links in the network.
    Furthermore, we show that the inverse of each submatrix can be computed in closed-form, thus significantly reducing the computational complexity.
    This complexity reduction is made possible by a keen observation of the special structure of the submatrix for each network entity.

\item Based on the decomposable structure of the Hessian matrix and the special second-order properties of the coefficient matrix of the MRFC problem, we further extend and generalize the matrix splitting idea of \cite{Jadbabaie09:DNewton,Wei10:DNewton} to the more complex MRFC problem.
    Also, we introduce a parameterized matrix splitting scheme so that the convergence performance of the iterative scheme for computing dual variables is tunable.

\item In addition to deriving closed-form expressions for the primal Newton direction and dual updates for each network entity, we also provide insights into the underlying networking interpretations of the proposed distributed Newton algorithm, as well as the connections to and differences from first-order approaches, thus further advancing our understanding of second-order approaches in network optimization theory.
\end{itemize}
To our knowledge, this paper is the first work that develops a distributed Newton algorithm for joint multi-path routing and flow control optimization.
Our work contributes to a new and exciting paradigm shift in cross-layer network design that is evolving from first-order to second-order methods.
We believe that, just as the intimate connection between the subgradient-based method and the ``back-pressure'' algorithm in the first-order paradigm, an interesting second-order version of the ``back-pressure'' algorithm may soon emerge, finding its roots in our proposed distributed Newton algorithm.

The remainder of this paper is organized as follows.
In Section~\ref{sec:Related}, we review some related work in the literature, putting our work in a comparative perspective.
Section~\ref{sec:model_formulation} introduces the network model and problem formulation.
Section~\ref{sec:Subgradient} briefly reviews the first-order decomposition approach and the subgradient algorithm to facilitate comparisons between first-order and second-order methods.
Section~\ref{sec:CNewton} provides some preliminary knowledge of the centralized Newton method and points out the difficulties in distributed implementations.
Section~\ref{sec:DNewton} is the key part of this paper, which develops the principal components of our proposed distributed Newton method.
Section~\ref{sec:numerical} provides some relevant numerical results, and Section~\ref{sec:conclusion} concludes this paper.

\section{Related Work} \label{sec:Related}
Early attempts at second-order methods for network optimization (centralized or distributed) date back to the 1980s \cite{Bertsekas83:PNewton,Klincewicz83:DNewton}.
In \cite{Bertsekas83:PNewton}, Bertsekas and Gafni employed a projected Newton method for multi-commodity flow problems.
The authors adapted a conjugate gradient approach \cite{Bazaraa_Sherali_Shetty_93:NLP} such that computing and storing the Hessian matrix is not needed.
However, a distributed implementation was not considered in this work.
In \cite{Klincewicz83:DNewton}, Klincewicz also proposed a distributed conjugate gradient direction method to solve a pure minimum cost flow routing problem where the network flows on each link are subject to more restrictive individual box-like constraints, as opposed to the more realistic sum capacity constraint over each link in this paper.
It was shown in \cite{Klincewicz83:DNewton} that feasible conjugate gradient directions can be computed distributedly using information exchange along a spanning tree.
However, the spanning tree computation still requires passing all information to a centralized node.
We remark that these conjugate gradient algorithms belong to the class of quasi-Newton methods that approximate the quadratically convergent Newton method by employing the notion of conjugate direction (see \cite{Bazaraa_Sherali_Shetty_93:NLP} for more details).
As a result, the convergence speed is relatively slower compared to pure Newton methods, although they could be simpler and more robust.
A more recent work on a distributed quasi-Newton method was reported in \cite{Bolognani10:DQNewton}, where Bolognani and Zampieri showed that the celebrated BFGS algorithm \cite{Bazaraa_Sherali_Shetty_93:NLP} can be decentralized to solve the optimal reactive power flow problem in smart-grids.

We also note that the early attempts in \cite{Bertsekas83:PNewton,Klincewicz83:DNewton} differ fundamentally from our work in that they rely on projecting gradients to find feasible search directions, while our algorithm belongs to the class of interior-point methods, which have been shown to more efficient \cite{Nesterov01:InteriorPoint}.
Indeed, most of the recent works in this area are based on the interior-point approach \cite{Zymnis07:DNewton,Jadbabaie09:DNewton,Bickson08:DNewton,Bickson09:DNewton,Wei10:DNewton}.
The first known interior-point based algorithm for a pure flow control problem (i.e., routes are fixed) was reported in \cite{Zymnis07:DNewton}, where Zymnis \etal proposed a centralized truncated-Newton primal-dual interior-point method.
For the same problem, Bickson \etal \cite{Bickson08:DNewton,Bickson09:DNewton} later developed a distributed algorithm based on Gaussian belief propagation technique, but without providing a provable guarantee for its convergence.
On the other hand, Jadbabaie \etal \cite{Jadbabaie09:DNewton} designed a distributed Newton method for solving a pure minimum cost routing problem (i.e., source flow rates are fixed), where a consensus-based local averaging scheme was used to compute the Newton direction.
Although convergence of the consensus-based scheme can be established by using spectral graph theory \cite{Chung-Graham94:GraphLaplacian}, we note that its convergence rate could potentially be slow in practice.
Our work is most related to \cite{Wei10:DNewton}, although \cite{Wei10:DNewton} studied the same pure flow control problem as in \cite{Zymnis07:DNewton,Bickson08:DNewton,Bickson09:DNewton}, while we consider a more complex joint multi-path routing and flow control problem.
As mentioned earlier, our dual update scheme is inspired by, and is a generalization of the matrix-splitting technique used in \cite{Wei10:DNewton}.
Although there exists some similarity in our matrix splitting approach with that in \cite{Wei10:DNewton}, we point out that due to a completely different network setting, showing the applicability of the matrix splitting technique in our problem is not straightforward and the resulting distributed algorithm is completely different.
Moreover, several interesting networking insights can be drawn from these new analyses and proofs.

Other than employing the aforementioned second-order methods, it is worth pointing out that the first-order subgradient method could also be modified to somewhat mimic the behavior of second-order methods.
Athuraliya and Low \cite{Athuraliya00:ScaledSubgradient} developed such a scaled subgradient method for the pure flow control problem.
Their basic idea is to use an appropriately scaled subgradient projection to approximate the diagonal terms of the Hessian matrix, while retaining their distributed nature.
Although the empirical convergence rate can be improved by using this approach, it does not achieve the same theoretical rate gain as second-order methods.

\section{Network Model and Problem Formulation} \label{sec:model_formulation}
We first introduce notation for matrices, vectors, and complex scalars used in this paper. We use
boldface to denote matrices and vectors.
For a matrix $\A$, $\A^{T}$ denotes the transpose of $\A$.
$\diag{\A_{1},\ldots,\A_{N}}$ represents the block diagonal matrix with matrices $\A_{1},\ldots,\A_{N}$ on its main diagonal.
$\diagv{\A}$ represents the vector containing the main diagonal entries of $\A$.
We let $(\A)_{ij}$ represent the entry in the $i$-th row and $j$-th column of $\A$.
We let $\I$ denote the identity matrix with dimension determined from the context.
$\A \succ 0$ represents that $\A$ is symmetric and positive definite (PD).
$\1$ and $\0$ denote vectors whose elements are all ones and zeros, respectively, where their dimensions are determined from the context.
$(\v)_{m}$ represents the $m$-th entry of any vector $\v$.
For a vector $\v$ and a matrix $\A$, $\v \geq \0$ and $\A \geq \0$ mean that $\v$ and $\A$ are element-wise nonnegative.

In this paper, a multi-hop network is represented by a directed graph, denoted by $\mathcal{G} = \{ \mathcal{N}, \mathcal{L} \}$, where $\mathcal{N}$ and $\mathcal{L}$ are the set of nodes and links, respectively.
We assume that $\mathcal{G}$ is connected.
The cardinalities of the sets $\mathcal{N}$ and $\mathcal{L}$ are $|\mathcal{N}|=N$ and $|\mathcal{L}|=L$, respectively.

We use the so-called {\em node-arc incidence matrix} (NAIM) \cite{Bazaraa_Jarvis_Sherali_90:LP}  $\A \in \mathbb{R}^{N \times L}$ to represent the network topology of $\mathcal{G}$.
The entry in $\A$ is defined as follows:
\begin{equation} \label{eqn_naim_a}
(\A)_{nl} =
\begin{cases}
1, & \text{if node $n$ is the transmitter node of link $l$}, \\
-1, & \text{if node $n$ is the receiving node of link $l$}, \\
0, & \text{otherwise.}
\end{cases}
\end{equation}

In the network, different source nodes send different data to their
intended destination nodes through {\em multi-path} and {\em multi-hop} routing.
Suppose that there is a total of $F$ sessions in the network, representing $F$ different commodities.
We denote the source and destination nodes of session $f$, $1 \leq f \leq F$ as $\src{f}$ and $\dst{f}$, respectively.
The source flow rate of session $f$ is denoted by a scalar $s_{f} \in \mathbb{R}_{+}$.
For session $f$, we use a {\em source--destination vector} vector $\b_{f} \in \mathbb{R}^{N}$ to represent the supply--demand relationship of session $f$.
More specifically, the entries in $\b_{f}$ are defined as follows:
\begin{equation} \label{eqn_supdmd_b}
(\b_{f})_{n} =
\begin{cases}
1, & \text{if node $n$ is a source node of session $f$}, \\
-1, & \text{if node $n$ is a destination node of session $f$}, \\
0, & \text{otherwise.}
\end{cases}
\end{equation}

\begin{figure}[t!]
\centering
\includegraphics[width=3.2in]{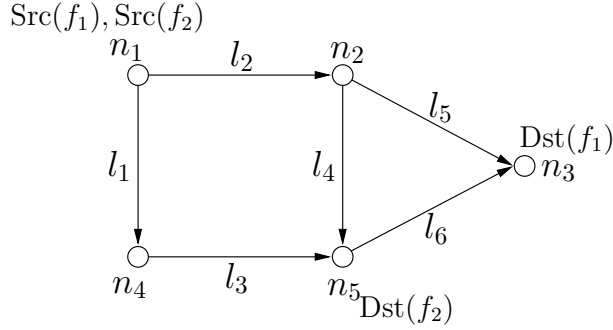}
\caption{A network example to illustrate the structure of $\A$ and $\b^{(f)}$.} \label{fig_net_ex}
\end{figure}
As an example, we use the 5-node 6-link network depicted in Fig.~\ref{fig_net_ex} to illustrate the structure of $\A$ and $\b^{(f)}$.
This network example will also be used throughout this paper to illustrate other concepts and their associated networking insights.
According to the definitions of $\A$ and $\b^{(f)}$, we have
\begin{equation*}
\A = \begin{tabular}{r|rrrrrr}
 & $l_{1}$ & $l_{2}$ & $l_{3}$ & $l_{4}$ & $l_{5}$ & $l_{6}$ \\
\hline $n_{1}$ & 1 & 1 & 0 & 0 & 0 & 0 \\
$n_{2}$ & 0 & $-1$ & 0 & 1 & 1 & 0 \\
$n_{3}$ & 0 & 0 & 0 & 0 & $-1$ & $-1$ \\
$n_{4}$ & $-1$ & 0 & 1 & 0 & 0 & 0 \\
$n_{5}$ & 0 & 0 & $-1$ & $-1$ & 0 & 1 \\
\end{tabular},
\quad
\b^{(1)} = \begin{tabular}{r|r}
 & $f_{1}$ \\
\hline $n_{1}$ & 1 \\
$n_{2}$ & 0 \\
$n_{3}$ & $-1$ \\
$n_{4}$ & 0 \\
$n_{5}$ & 0 \\
\end{tabular},
\quad \text{and} \quad
\b^{(2)} = \begin{tabular}{r|r}
 & $f_{2}$ \\
\hline $n_{1}$ & 1 \\
$n_{2}$ & 0 \\
$n_{3}$ & 0 \\
$n_{4}$ & 0 \\
$n_{5}$ & $-1$ \\
\end{tabular}.
\end{equation*}
It can be seen that a distinct feature of $\A$ and $\b^{(f)}$ is that each column has exactly two non-zero entries: a ``$1$'' and ``$-1$.''

For every link $l$, we let $x_{l}^{(f)} \geq 0$ represent the flow amount of session $f$ on link $l$.
We assume that the network is a flow-balanced system, i.e., the following flow balance constraints hold at each node:
\begin{equation}
\label{eqn_flowbalance_src}
\sum_{l \in \Out{n}} x_{l}^{(f)} - \sum_{l \in \In{n}} x_{l}^{(f)} = s_{f}, \quad \text{if $n = \src{f}$},
\end{equation}

\begin{equation} \label{eqn_flowblance_nonsrc}
\sum_{l \in \Out{n}} x_{l}^{(f)} = \sum_{l \in \In{n}} x_{l}^{(f)}, \quad \text{if $n \ne \src{f},\dst{f}$,}
\end{equation}

\begin{equation} \label{eqn_flowbalance_dst}
\sum_{l \in \In{n}} x_{l}^{(f)} - \sum_{l \in \Out{n}} x_{l}^{(f)} = s_{f}, \quad \text{if $n = \dst{f}$},
\end{equation}
where $\Out{n}$ and $\In{n}$ represent the sets of outgoing and incoming links at node $n$, respectively.
We define $\x^{(f)} \triangleq [ x_{1}^{(f)}, \ldots, x_{L}^{(f)} ]^{T} \in \mathbb{R}^{L}$ as the {\em routing vector} for session $f$ across all links.
Using the notation $\A$, $\b_{f}$, and $\x^{(f)}$, the flow balance constraints above can be compactly written as
\begin{equation} \label{eqn_flowbalance}
\A \x^{(f)} - s_{f}\b_{f} = \0, \quad \forall f=1,2,\ldots,F.
\end{equation}
Moreover, upon taking a closer look at the linear equation system in (\ref{eqn_flowbalance}), it is easy to see \cite{Bazaraa_Jarvis_Sherali_90:LP} that the coefficient matrix $\A$ is not full row rank (because all columns sum up to zero).
To eliminate the redundant rows in $\A$, we let $\A^{(f)} \in \mathbb{R}^{(N-1)\times L}$ be obtained by deleting from $\A$ the row corresponding to the node $\dst{f}$.
It is easy to verify that $\A^{(f)}$ is of full row rank \cite{Bazaraa_Jarvis_Sherali_90:LP}.
Also, we let $\bt^{(f)} \in \mathbb{R}^{N-1}$ be obtained by deleting from $\b^{(f)}$ the entry corresponding to the node $\dst{f}$.
\begin{figure}[t!]
    \begin{minipage}[t]{0.45\linewidth}
        \centering
        \includegraphics[width=2.6in]{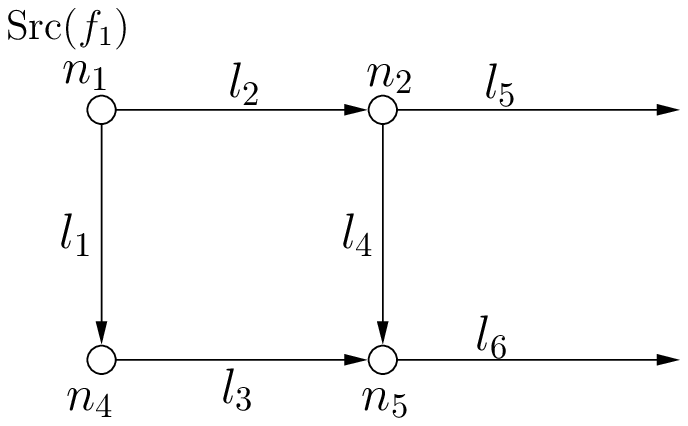}
        \caption{$\A^{(1)}$ and $\b^{(1)}$ are obtained by deleting the destination node of $f_{1}$, i.e., node 3.} \label{fig_net_ex_f1}
    \end{minipage}%
    \hspace{.1in}
    \begin{minipage}[t]{0.45\linewidth}
        \centering
        \includegraphics[width=2.6in]{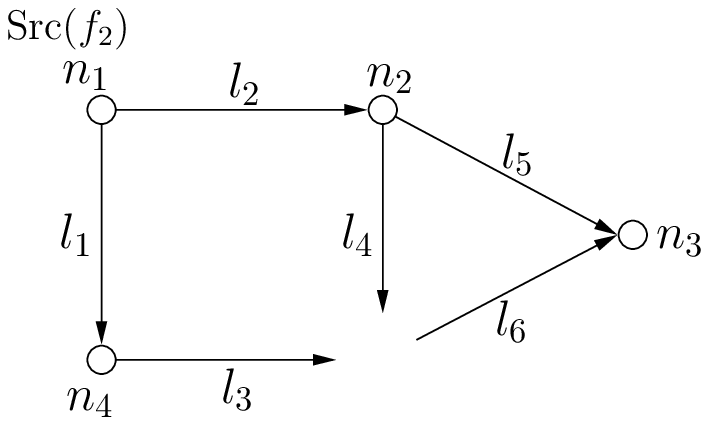}
        \caption{$\A^{(2)}$ and $\b^{(2)}$ are obtained by deleting the destination node of $f_{2}$, i.e., node 5.} \label{fig_net_ex_f2}
    \end{minipage}
\end{figure}
For example, for the network in Fig.~\ref{fig_net_ex}, by deleting the third and fifth rows (i.e., corresponding to deleting the nodes 3 and 5, see Figs.~\ref{fig_net_ex_f1} and~\ref{fig_net_ex_f2}), we can write $\A^{(1)}$, $\A^{(2)}$, $\bt^{(1)}$, and $\bt^{(2)}$ as follows:
\begin{equation*}
\A^{(1)} = \begin{tabular}{r|rrrrrr}
 & $l_{1}$ & $l_{2}$ & $l_{3}$ & $l_{4}$ & $l_{5}$ & $l_{6}$ \\
\hline $n_{1}$ & 1 & 1 & 0 & 0 & 0 & 0 \\
$n_{2}$ & 0 & $-1$ & 0 & 1 & 1 & 0 \\
$n_{4}$ & $-1$ & 0 & 1 & 0 & 0 & 0 \\
$n_{5}$ & 0 & 0 & $-1$ & $-1$ & 0 & 1 \\
\end{tabular},
\quad \quad
\bt^{(1)} = \begin{tabular}{r|r}
 & $f_{1}$ \\
\hline $n_{1}$ & 1 \\
$n_{2}$ & 0 \\
$n_{4}$ & 0 \\
$n_{5}$ & 0 \\
\end{tabular},
\end{equation*}
\begin{equation*}
\A^{(2)} = \begin{tabular}{r|rrrrrr}
 & $l_{1}$ & $l_{2}$ & $l_{3}$ & $l_{4}$ & $l_{5}$ & $l_{6}$ \\
\hline $n_{1}$ & 1 & 1 & 0 & 0 & 0 & 0 \\
$n_{2}$ & 0 & $-1$ & 0 & 1 & 1 & 0 \\
$n_{3}$ & 0 & 0 & 0 & 0 & $-1$ & $-1$ \\
$n_{4}$ & $-1$ & 0 & 1 & 0 & 0 & 0 \\
\end{tabular},
\quad \quad
\bt^{(2)} = \begin{tabular}{r|r}
 & $f_{2}$ \\
\hline $n_{1}$ & 1 \\
$n_{2}$ & 0 \\
$n_{3}$ & 0 \\
$n_{4}$ & 0 \\
\end{tabular}.
\end{equation*}

We assume that each link in the network is capacitated.
The capacity of link $l$, denoted by $C_{l}$, is assume to be fixed, which models conventional wireline networks or wireless networks with static channel and fixed transmit power.
Since the total network flow traversing a link cannot exceed the link's capacity limit, we have
\begin{equation*} \label{eqn_link_limit}
\sum_{f=1}^{F} x_{l}^{(f)} \leq C_{l}, \quad \forall l = 1, \ldots, L.
\end{equation*}

We associate a utility function $U_{f}: \mathbb{R}_{+} \rightarrow \mathbb{R}$ with each session $f$, i.e., $U_{f}(s_{f})$ denotes the utility of session $f$ as a function of the session rate $s_{f}$.
We assume that the utility functions are additive so that the overall network utility is given by $\sum_{f=1}^{F} U_{f}(s_{f})$.
We also assume that the utility functions $U_{f}$ are strictly concave, monotonically increasing, twice continuously differentiable, and reversely self-concordant (see \cite{Boyd_Vandenberghe04:Cnvx_Opt} for the definition of self-concordance).
Our objective is to maximize the sum of utilities of all sessions.
Putting together the routing and flow control constraints described earlier, we can formulate the joint multi-hop routing and flow control optimization (MRFC) problem as follows:
\begin{align*}
& \text{{\bf MRFC: }} \\
& \text{Maximize} && \hspace{-.8in} \sum_{f=1}^{F} U_{f}(s_{f}) & \\
& \text{subject to} && \hspace{-.8in} \A^{(f)} \x^{(f)} - s_{f} \bt^{(f)} = \0, && \hspace{-1in} \forall f=1,\ldots,F \\
& && \hspace{-.8in} \sum_{f=1}^{F} x_{l}^{(f)} \leq C_{l}, && \hspace{-1in} \forall l=1,\ldots,L \\
& && \hspace{-.8in} x_{l}^{(f)} \geq 0, \,\, \forall f,l; \quad s_{f} \geq 0, \,\, \forall f.
\end{align*}

Note that, in MRFC, the objective function is concave and all constraints are linear.
Hence, this problem is a convex program and can be solved by using standard convex programming methods.
Moreover, due to a decomposable structure in the dual domain, it is well-known that the MRFC problem can also be solved in a distributed fashion based on a Lagrangian dual decomposition and subgradient optimization framework.
In the next section, we will briefly review the dual subgradient method, which will later be compared with our proposed Newton method.

\section{Dual Subgradient Method for Solving MRFC: A Quick Overview} \label{sec:Subgradient}
Since MRFC is a linearly constrained convex program, it can be equivalently solved in its dual domain because of a zero duality gap.
To solve the MRFC problem in its dual domain, we first slightly modify the first constraint in MRFC as an inequality constraint $\A^{(f)} \x^{(f)} - s_{f} \bt^{(f)} \geq \0$.
This modification does not affect the solution at optimality and can be interpreted from a network stability perspective (i.e., total service rate at each node is no less than the total arrival rate).
Then, by associating a dual variable $u_{n}^{(f)} \geq 0$ for all $n,f$ and rearranging terms in the Lagrangian, it can be shown that the dual function can be written as
\begin{equation*}
\Theta(\u) = \sum_{f=1}^{F} \Theta_{\mathrm{FC}}\Big(u_{\src{f}}^{(f)}\Big) + \sum_{l=1}^{L} \Theta_{\mathrm{R}}\Big(u_{\tx{l}}^{(f)},u_{\rx{l}}^{(f)}\Big),
\end{equation*}
where $\Theta_{\mathrm{FC}}\Big(u_{\src{f}}^{(f)}\Big)$ and $\Theta_{\mathrm{R}}\Big(u_{\tx{l}}^{(f)},u_{\rx{l}}^{(f)}\Big)$ are respectively corresponding to the flow-control subproblem at node $\src{f}$ (transport layer) and the routing subproblem at each link $l$ (network layer):
\begin{equation} \label{eqn_transp_subprob}
\Theta_{\mathrm{FC}}\Big(u_{\src{f}}^{(f)}\Big) \triangleq
\max\left\{ U_{f}(s_{f}) - u_{\src{f}}^{(f)} s_{f} \left| s_{f}\geq 0 \right. \right\},
\end{equation}

\begin{equation} \label{eqn_net_subprob}
\Theta_{\mathrm{R}}\Big(u_{\tx{l}}^{(f)},u_{\rx{l}}^{(f)}\Big) \triangleq
\max\Big\{ \sum_{f=1}^{F} \big(u_{\tx{l}}^{(f)}-u_{\rx{l}}^{(f)}\big) x_{l}^{(f)} \Big| x_{l}^{(f)}\geq 0, \text{and} \sum_{f=1}^{F} x_{l}^{(f)} \leq C_{l} \,\, \forall f \Big. \Big\}.
\end{equation}
The dual problem can be written as
\begin{equation}
\begin{array}{rl}
\text{Minimize} & \Theta(\u) \\
\text{subject to} & \u \geq \0.
\end{array}
\end{equation}
Due to this separable structure, the dual function $\Theta(\u)$ can be evaluated by computing $\Theta_{\mathrm{FC}}\Big(u_{\src{f}}^{(f)}\Big)$ and $\Theta_{\mathrm{R}}\Big(u_{\tx{l}}^{(f)},u_{\rx{l}}^{(f)}\Big)$ for each source node and each link, respectively.
The optimal dual variables $\u^{*}$ can be iteratively computed by using the subgradient method as follows:
\begin{equation} \label{eqn_sg_dual_update}
u_{n}^{(f)}(k+1) = \max\{u_{n}^{(f)}(k) - \pi^{k} d_{n}^{(f)}(k), 0\}, \quad \forall n,f,
\end{equation}
where $\pi^{k} > 0$ is a step size chosen at the $k$-th iteration and $d_{n}^{(f)}(k)$ is a subgradient at the $k$-th iteration, which can be computed as
\begin{equation} \label{eqn_subgradient}
d_{n}^{(f)}(k) =
\begin{cases}
\sum_{l\in \Out{n}} x_{l}^{(f)}(k) - \sum_{l \in \In{n}} x_{l}^{(f)}(k), & \text{if } n \ne \src{f}, \\
\sum_{l\in \Out{n}} x_{l}^{(f)}(k) - s_{f}(k), & \text{if } n = \src{f}.
\end{cases}
\end{equation}
It can be seen from (\ref{eqn_subgradient}) that the subgradient $d_{n}^{(f)}(k)$ can also be computed at each node in a decentralized fashion.

There are several interesting networking insights in the subgradient-based first-order method.
First, the dual variables $u_{n}^{(f)}$ can be interpreted as the price charged to session $f$ by node $n$.
For example, if the subgradient component $d_{n}^{(f)}(k)=\sum_{l\in \Out{n}} x_{l}^{(f)}(k) - \sum_{l \in \In{n}} x_{l}^{(f)}(k) < 0$, i.e., the stability condition is violated, then the price $w_{n}^{(f)}$ will increase in the $(k+1)$-st iteration, thus discouraging session $f$ from passing through node $n$.
Second, it can be seen that by dividing the step size $\pi^{k}$ on both sides of (\ref{eqn_sg_dual_update}) and letting $Q_{n}^{(f)}(k) \triangleq u_{n}^{(f)}/\pi^{k}$, we have $Q_{n}^{(f)}(k+1) = \max\{Q_{n}^{(f)}(k) - \sum_{l\in \Out{n}} x_{l}^{(f)}(k) + \sum_{l \in \In{n}} x_{l}^{(f)}(k), 0\}$ if $n\ne \src{f}$ or $Q_{n}^{(f)}(k+1) = \max\{Q_{n}^{(f)}(k) - \sum_{l\in \Out{n}} x_{l}^{(f)}(k) + s_{f}, 0\}$ if $n=\src{f}$.
This is exactly the queue length evolution of session $f$ at node $n$.
Thus, the dual variables and queue lengths are intimately related (differ by a scaling factor).
Lastly, it is easy to see that the knapsack-type subproblem $\Theta_{\mathrm{R}}\Big(u_{\tx{l}}^{(f)},u_{\rx{l}}^{(f)}\Big)$ admits a trivial solution: for the given link $l$, pick a session that has the largest $\big(u_{\tx{l}}^{(f)}-u_{\rx{l}}^{(f)}\big)$-value quantity, say $f^{*}$, and let $f^{*}$ use up the link capacity $C_{l}$.
This is exactly the same strategy used in the celebrated ``back-pressure'' algorithm that was first discovered in \cite{Tassiulas92:BackPressure}, even though its throughput-optimality was established using tools in control theory.

However, despite its simplicity and interesting networking interpenetrations, the subgradient method usually does not work well in practice due to its slow rate of convergence and sensitivity to step-size selection.

\section{Centralized Newton Method: A Primer} \label{sec:CNewton}
In this section, we provide some preliminary discussion on using the conventional Newton method to solve Problem MRFC, along with a problem reformulation and an analysis of the challenges in developing distributed algorithms.

\subsection{Problem Reformulation} \label{sec:reformulate}
To facilitate the development of a distributed Newton method for solving MRFC, we need to reformulate MRFC into a form that only has equality constraints so that the Newton method can be readily applied.
Following the standard approach used for interior point methods \cite{Nesterov01:InteriorPoint}, we employ a logarithmic barrier function to represent each inequality constraint including non-negativity restrictions, and we accommodate this within the objective function.
The augmented objective function is rewritten as follows:
\begin{equation*}
f(\y) = -t\sum_{f=1}^{F} U_{f}(s_{f}) -
\sum_{l=1}^{L} \log \bigg( C_{l} - \sum_{f=1}^{F} x_{l}^{(f)} \bigg) - \sum_{f=1}^{F} \log(s_{f}) - \sum_{l=1}^{L} \sum_{f=1}^{F} \log(x_{l}^{(f)}),
\end{equation*}
where $\y \triangleq \Big[s_{1} \cdots s_{F}, (\x^{(1)})^{T}, \cdots, (\x^{(F)})^{T} \Big]^{T} \in \mathbb{R}^{(L+1)F}$, and where $t>0$ is a parameter that permits us to track the central path in the interior point method as $t\rightarrow \infty$ \cite{Boyd_Vandenberghe04:Cnvx_Opt}.
Furthermore, we denote
$
\M \triangleq
\mtrx{{ccc|ccc}
\bt^{(1)} & & & -\A^{(1)} & & \\
 & \ddots & & & \ddots & \\
 & & \bt^{(F)} & & & -\A^{(F)}
} \in \mathbb{R}^{(N-1)F \times (L+1)F}
$.
Clearly, $\M$ is of full row rank because of its block diagonal structure with each block being of full row rank.
With this notation, we can rewrite the revised Problem MRFC as follows:
\begin{equation}
\begin{array}{rl}
 & \hspace{-.8in} \text{{\bf R-MRFC: }}  \\
\text{Minimize} & f(\y) \\
\text{subject to} & \M \y = \0.
\end{array}
\end{equation}

Observe that the approximation accuracy of R-MRFC can be controlled by $t$:
as $t$ increases, the first term in $f(\y)$ dominates the barrier functions and R-MRFC becomes a better approximation of MRFC.
In fact, it can be shown that the approximation error is bounded by $K/t$ \cite{Boyd_Vandenberghe04:Cnvx_Opt}, where $K$ is some constant depending on problem specifics.
In other words, as $t\rightarrow \infty$, the solution of R-MRFC converges to that for the original problem.
However, one may suspect that, as $t$ increases, solving R-MRFC may become more difficult due to numerical instability.
Interestingly, this is not the case as long as $f(\y)$ satisfies the so-called self-concordance conditions, which has been guaranteed because $-U_{f}(\cdot)$ and all barrier functions are self-concordant (we refer readers to \cite{Nesterov01:InteriorPoint,Boyd_Vandenberghe04:Cnvx_Opt} for more detailed discussions on self-concordance).

\subsection{Centralized Newton Method for the Reformulated Problem}
Starting from an initial feasible solution $\y^0$, the centralized Newton method searches for an optimal solution using the following iterative rule:
\begin{equation} \label{eqn_newton_iterate}
\y^{k+1} = \y^{k} + \pi^{k} \Dy^{k},
\end{equation}
where $\pi^{k} > 0$ is a positive step-size.
In (\ref{eqn_newton_iterate}), $\Dy^{k}$ denotes the Newton direction, which is the solution to the following linear equation system (obtained by deriving the Karash-Kuhn-Tucker (KKT) system of the second-order approximation of $f(\y)$ \cite{Bazaraa_Sherali_Shetty_93:NLP,Boyd_Vandenberghe04:Cnvx_Opt}):
\begin{equation} \label{eqn_newton_linearsys}
\mtrx{{cc}
\Hc_{k} & \M^{T} \\
\M & \0
}
\mtrx{{c}
\Dy^{k} \\
\w^{k}
}
=
-\mtrx{{c}
\nabla f(\y^{k}) \\
\0
},
\end{equation}
where $\Hc_{k} \triangleq \nabla^{2} f(\y^{k}) \in \mathbb{R}^{(L+1)F \times (L+1)F}$ is the Hessian matrix of $f(\y)$ at $\y^{k}$, and the vector $\w^{k} \in \mathbb{R}^{(N-1)F}$ contains the dual variables for the flow balance constraints $\M \y = \0$ at the $k$-th iteration.

It can be easily verified that the coefficient matrix of the linear equation in (\ref{eqn_newton_linearsys}) is nonsingular.
Therefore, the primal direction $\Delta \y^{k}$ and the dual variables $\w^{k}$ can be uniquely determined by solving (\ref{eqn_newton_linearsys}).
However, solving $\Delta \y^{k}$ and $\w^{k}$ simultaneously via (\ref{eqn_newton_linearsys}) requires global information.
A key step towards solving (\ref{eqn_newton_linearsys}) in a decentralized manner is to rewrite it as follows:
\begin{align}
\label{eqn_newton_linsys_1} & \Dy^{k} = -\Hc_{k}^{-1} (\nabla f(\y^{k}) + \M^{T} \w^{k}), \\
\label{eqn_newton_linsys_2} & (\M \Hc_{k}^{-1} \M^{T}) \w^{k} = -\M \Hc_{k}^{-1} \nabla f(\y^{k}).
\end{align}
Thus, given $\y^{k}$, we can solve for $\w^{k}$ from (\ref{eqn_newton_linsys_2}), and hence, solve for $\Delta \y^{k}$ from (\ref{eqn_newton_linsys_1}), which can thus be used in (\ref{eqn_newton_iterate}) along with an appropriate step-size $\pi^{k}$.
However, as we shall see in Section~\ref{sec:DNewton}, the inverses of $\Hc_{k}$ and $(\M \Hc_{k}^{-1} \M^{T})$ still require global information.
This is unlike the pure flow control problem in \cite{Wei10:DNewton} and the pure minimum cost routing problem in \cite{Jadbabaie09:DNewton}, where the Hessian matrix $\Hc_{k}$ is diagonal and can be readily computed using local information.
Therefore, in the next section, our goal is to further reformulate R-MRFC so that we can design an iterative scheme to compute $\Delta \y^{k}$ and $\w^{k}$ in a distributed fashion.

\section{Distributed Newton Method} \label{sec:DNewton}
In this section, we will present the key components of our proposed distributed Newton method.
We will first further reformulate Problem R-MRFC in Section~\ref{sec:MRFC_Reform}, following which we will introduce some basic second-order structural properties of the reformulated problem in Section~\ref{sec:Basic_2nd_Prop}.
The distributed computations of the primal Newton direction and the dual variables will be presented in Sections~\ref{sec:Primal_Newton_Dir} and~\ref{sec:Dual_Variable_Compu}, respectively.
Finally, in Section~\ref{sec:DNewton_implement}, we shall discuss some other implementation issues (i.e., information exchange mechanism, initialization, stopping criterion, and step-size selection).

\subsection{Problem Rearrangement} \label{sec:MRFC_Reform}
The first major hurdle in decentralizing the Newton method is the coupled structure in the Hessian matrix of the MRFC problem.
To see this, we start by evaluating the first and second partial derivatives of $f(\y)$.
Noting that $f(\y)$ is separable with respect to each flow $f$ and link $l$, the only non-zero partial derivatives are:
\begin{align*}
& \frac{\partial f(\y^{k})}{\partial s_{f}} = -tU'_{f}(s_{f})-\frac{1}{s_{f}}, &&\forall f, & &\frac{\partial f(\y^{k})}{\partial x_{l}^{(f)}} = \frac{1}{C_{l}-\sum_{f'=1}^{F} x_{l}^{(f')}} - \frac{1}{x_{l}^{(f)}}, &&\forall l, f, \\
& \frac{\partial^2 f(\y^{k})}{\partial (s_{f})^{2}} = -tU''_{f}(s_{f})+\frac{1}{(s_{f})^{2}}, &&\forall f, & &\frac{\partial^{2} f(\y^{k})}{\partial (x_{l}^{(f)})^{2}} = \frac{1}{(C_{l}-\sum_{f'=1}^{F} x_{l}^{(f')})^{2}} + \frac{1}{(x_{l}^{(f)})^{2}}, &&\forall l, f,\\
& && & &\frac{\partial^{2} f(\y^{k})}{\partial x_{l}^{(f_{1})} \partial x_{l}^{(f_{2})}} = \frac{1}{(C_{l}-\sum_{f'=1}^{F} x_{l}^{(f')})^{2}}, &&\forall l, f_{1} \ne f_{2}.
\end{align*}
For convenience, we use $\delta_{l} \triangleq C_{l}-\sum_{f=1}^{F} x_{l}^{(f)}$ to represent the {\em unused link capacity} of link $l$, which will occur frequently in the rest of the paper.
We also define three types of diagonal matrices:
\begin{align*}
&\D_{0} = \diag{-tU''_{f}(s_{f}) + \frac{1}{(s_{f})^{2}}, f=1,\ldots,F},\\
&\D_{1}^{(f)} = \diag{ \frac{1}{\delta_{l}^{2}} + \frac{1}{(x_{l}^{(f)})^{2}}, l=1,\ldots,L},\\
&\D_{2} = \diag{ \frac{1}{\delta_{l}^{2}}, l=1,\ldots,L}.
\end{align*}
Then, it can be verified that $\Hc_{k}$ has the following structure:
\begin{equation*}
\Hc_{k} \triangleq \nabla^{2} f(\y^{k}) =
\mtrx{{c|cccc}
\D_{0} & & & & \\
\hline & \D_{1}^{(1)} & \D_{2} & \cdots & \D_{2} \\
 & \D_{2} & \D_{1}^{(2)} & \cdots & \D_{2} \\
 & \vdots & \vdots & \ddots & \vdots \\
 & \D_{2} & \D_{2} & \cdots & \D_{1}^{(F)} \\
}.
\end{equation*}
Due to this non-diagonal and coupled structure with respect to the different sessions, $\Hc_{k}^{-1}$ cannot be computed separably for each source node and each link.
Fortunately, this problem can be tackled by reformulating R-MRFC as described next.

First, from the partial derivatives computation, we note that $f(\y)$ is separable with respect to each link.
This prompts us to rearrange $\y$ and $\M$ based on links as follows:
\begin{equation*}
\yt \triangleq \Big[ s_{1} \cdots s_{F} \big| x_{1}^{(1)} \cdots x_{1}^{(F)} \big| \cdots \cdots \big| x_{L}^{(1)} \cdots x_{L}^{(F)} \Big]^{T} \in \mathbb{R}^{(L+1)F}, \quad \text{and}
\end{equation*}

\begin{equation*}
\Mt = \mtrx{{cccc} \Bt & \A_{1} & \cdots & \A_{L} },
\end{equation*}
where
\begin{equation*}
\Bt \triangleq \mtrx{{ccc}
\bt^{(1)} & & \\
 & \ddots & \\
 & & \bt^{(F)}
}
\quad \text{and} \quad
\A_{l} \triangleq \mtrx{{ccc}
-\a_{l}^{(1)} & & \\
 & \ddots & \\
 & & -\a_{l}^{(F)}
},
\,\, l=1,\ldots,L,
\end{equation*}
and where in the definition of $\A_{l}$, the vector $\a_{l}^{(f)}$ is the $l$-th column in the matrix $\A^{(f)}$, i.e., $\A^{(f)} = \mtrx{{cccc} \a_{1}^{(f)}, \a_{2}^{(f)}, \cdots, \a_{L}^{(f)}}$.
For example, for the network in Fig.~\ref{fig_net_ex}, $\a_{2}^{(1)}$ and $\a_{4}^{(2)}$ can be written as
\begin{equation*}
\a_{2}^{(1)} = \begin{tabular}{r|r}
 & $l_{2}$ \\
\hline $n_{1}$ & 1 \\
$n_{2}$ & $-1$ \\
$n_{4}$ & 0 \\
$n_{5}$ & 0
\end{tabular},
\quad
\a_{4}^{(2)} = \begin{tabular}{r|r}
 & $l_{4}$ \\
\hline $n_{1}$ & 0 \\
$n_{2}$ & 1 \\
$n_{3}$ & 0 \\
$n_{4}$ & 0
\end{tabular}.
\end{equation*}
As a result, R-MRFC can be equivalently re-written as follows:
\begin{equation}
\begin{array}{rl}
 & \hspace{-.8in} \text{{\bf R2-MRFC: }} \\
\text{Minimize} & f(\yt) \\
\text{subject to} & \Mt \yt = \0.
\end{array}
\end{equation}
By the same token as in the previous section, the Newton direction of R2-MRFC is the solution to the following linear equation system:
\begin{equation} \label{eqn_newton_linearsys2}
\mtrx{{cc}
\tHc_{k} & \Mt^{T} \\
\Mt & \0
}
\mtrx{{c}
\Dyt^{k} \\
\wt^{k}
}
=
-\mtrx{{c}
\nabla f(\yt^{k}) \\
\0
},
\end{equation}
where $\wt^{k}$ represents the dual variables for the flow balance constraint $\Mt \yt = \0$.
Here, the entries in $\wt^{k}$ are arranged as $[(\wt_{k}^{(1)})^{T}, \ldots (\wt_{k}^{(F)})^{T}]^{T}$, where $\wt_{k}^{(f)}$ is in the form of
\begin{equation} \label{eqn_dual_arrange}
\wt_{k}^{(f)} \triangleq \big[\widetilde{w}_{1}^{(f)},\ldots, \widetilde{w}_{\dst{f}-1}^{(f)}, \widetilde{w}_{\dst{f}+1}^{(f)}, \ldots, \widetilde{w}_{N}^{(f)} \big]^{T} \in \mathbb{R}^{N-1}.
\end{equation}
Note that in (\ref{eqn_dual_arrange}), we have dropped the iteration index $k$ within $[\cdot]$for notational simplicity.
For the same reason, in the rest of the paper, the iteration index $k$ will be dropped whenever such an omission does not cause confusion.
Also, we let $\widetilde{w}_{\dst{f}}^{(f)} \equiv 0$, for all $f$.
As we shall see later, this helps simplify the closed-form expressions in Theorem~\ref{thm:primal_newton_direct}.
More detailed discussions on the physical meaning of the dual variables $\wt^{k}$ will also be provided in Section~\ref{sec:Primal_Newton_Dir}.

\subsection{Basic Second-Order Properties of $\a_{l}^{(f)}$ and $\bt^{(f)}$} \label{sec:Basic_2nd_Prop}
In formulating Problem R2-MRFC, we have introduced two new vectors, namely, $\a_{l}^{(f)}$ and $\bt^{(f)}$.
Here, we will first study some of their basic second-order properties, which will be used extensively later in designing a distributed Newton method.
Most of these properties can be verified using simple matrix computations based on the definitions of $\a_{l}^{(f)}$ and $\bt^{(f)}$.
Thus, we omit the formal proofs of these properties.
First, we have the following basic second-order property for $\bt{(f)}$:
\begin{lem} \label{lem:btfbtft}
The rank-one matrix $\bt^{(f)} (\bt^{(f)})^{T}$ has the following structure:
\begin{equation}
\Big(\bt^{(f)} (\bt^{(f)})^{T} \Big)_{ij} =
\begin{cases}
1, & \text{if $i=j$ and $(\bt^{(f)})_{i}$ corresponds to $\tx{l}$}. \\
0, & \text{otherwise}.
\end{cases}
\end{equation}
\end{lem}

\noindent For $\a_{l}^{(f)}$, we have the following two second-order properties.
\begin{lem} \label{lem:alfalft}
The rank-one matrix $\a_{l}^{(f)} (\a_{l}^{(f)})^{T}$ has the following structure:
\begin{itemize}
\item Case 1: If none of link $l$'s two end points is the destination node of flow $f$, i.e., $\tx{l},\rx{l}\ne \dst{f}$, then $\a_{l}^{(f)} (\a_{l}^{(f)})^{T}$ has four non-zero entries, where
    \begin{equation*}
    \Big(\a_{l}^{(f)} (\a_{l}^{(f)})^{T} \Big)_{ij} =
    \begin{cases}
    1, & \text{if $i=j$, $(\a_{l}^{(f)})_{i}$ corresponds to $\tx{l}$ or $\rx{l}$}, \\
    -1, & \text{if $i \ne j$, $(\a_{l}^{(f)})_{i}$ corresponds to $\tx{l}$ and $(\a_{l}^{(f)})_{j}$ corresponds to $\rx{l}$}, \\
      & \text{or $(\a_{l}^{(f)})_{j}$ corresponds to $\tx{l}$ and $(\a_{l}^{(f)})_{i}$ corresponds to $\rx{l}$}, \\
    0, & \text{otherwise};
    \end{cases}
    \end{equation*}

\item Case 2: If link $l$'s receiving node is the destination node of flow $f$, i.e., $\rx{l} = \dst{f}$ then $\a_{l}^{(f)} (\a_{l}^{(f)})^{T}$ has one non-zero element, where
    \begin{equation}
    \Big( \a_{l}^{(f)} (\a_{l}^{(f)})^{T} \Big)_{ij} =
    \begin{cases}
    1, & \text{if $i=j$ and $(\a_{l}^{(f)})_{i}$ corresponds to node $\tx{l}$}, \\
    0, & \text{otherwise}.
    \end{cases}
    \end{equation}
\end{itemize}
\end{lem}

\begin{lem} \label{lem:alf1alf2t}
The rank-one matrix $\a_{l}^{(f_{1})} (\a_{l}^{(f_{2})})^{T}$ has the following structure:
\begin{itemize}
\item Case 1: If none of link $l$'s two end points is the destination node of either flow $f_{1}$ or flow $f_{2}$, i.e., $\tx{l},\rx{l}\ne \dst{f_{1}}$ and $\tx{l},\rx{l}\ne \dst{f_{2}}$, then $\a_{l}^{(f_{1})} (\a_{l}^{(f_{2})})^{T}$ has four non-zero entries, where
    \begin{equation*}
    \Big(\a_{l}^{(f_{1})} (\a_{l}^{(f_{2})})^{T} \Big)_{ij} =
    \begin{cases}
    1, & \text{if $(\a_{l}^{(f_{1})})_{i}$ and $(\a_{l}^{(f_{2})})_{j}$ both correspond to $\tx{l}$}, \\
      & \text{or $(\a_{l}^{(f_{1})})_{j}$ and $(\a_{l}^{(f_{2})})_{i}$ both correspond to $\tx{l}$}, \\
    -1, & \text{if $(\a_{l}^{(f_{1})})_{i}$ corresponds to $\tx{l}$ and $(\a_{l}^{(f_{2})})_{j}$ corresponds to $\rx{l}$}, \\
      & \text{or $(\a_{l}^{(f_{1})})_{j}$ corresponds to $\tx{l}$ and $(\a_{l}^{(f_{2})})_{i}$ corresponds to $\rx{l}$}, \\
    0, & \text{otherwise};
    \end{cases}
    \end{equation*}

\item Case 2: If link $l$'s receiving node is the destination node of either flow $f_{1}$ or flow $f_{2}$, i.e., $\rx{l} = \dst{f_{1}}$ or $\rx{l} = \dst{f_{2}}$, then $\a_{l}^{(f_{1})} (\a_{l}^{(f_{2})})^{T}$ has two non-zero elements, where
    \begin{equation*}
    \Big( \a_{l}^{(f_{1})} (\a_{l}^{(f_{2})})^{T} \Big)_{ij} =
    \begin{cases}
    1, & \text{if $(\a_{l}^{(f_{1})})_{i}$ and $(\a_{l}^{(f_{2})})_{j}$ both correspond to $\tx{l}$}, \\
    -1, & \text{if $(\a_{l}^{(f_{1})})_{i}$ corresponds to $\tx{l}$ and $(\a_{l}^{(f_{2})})_{j}$ corresponds to $\rx{l}$}, \\
     & \text{or $(\a_{l}^{(f_{1})})_{i}$ corresponds to $\rx{l}$ and $(\a_{l}^{(f_{2})})_{j}$ corresponds to $\tx{l}$}, \\
    0, & \text{otherwise}.
    \end{cases}
    \end{equation*}

\item Case 3: If the two end nodes of link $l$ are respectively the destination nodes of flow $f_{1}$ and flow $f_{2}$, then $\a_{l}^{(f_{1})} (\a_{l}^{(f_{2})})^{T}$ has one non-zero element, where
    \begin{equation*}
    \Big( \a_{l}^{(f_{1})} (\a_{l}^{(f_{2})})^{T} \Big)_{ij} =
    \begin{cases}
    -1, & \text{if $(\a_{l}^{(f_{1})})_{i}$ corresponds to $\tx{l}$ and $(\a_{l}^{(f_{2})})_{j}$ corresponds to $\rx{l}$}, \\
     & \text{or $(\a_{l}^{(f_{2})})_{i}$ corresponds to $\rx{l}$ and $(\a_{l}^{(f_{2})})_{j}$ corresponds to $\tx{l}$}, \\
    0, & \text{otherwise}.
    \end{cases}
    \end{equation*}
\end{itemize}
\end{lem}

For example, for the network in Fig.~\ref{fig_net_ex}, we have
\begin{equation*}
\a_{1}^{(1)} (\a_{1}^{(1)})^{T} =
\begin{tabular}{r|cccc}
 & $n_{1}$ & $n_{2}$ & $n_{4}$ & $n_{5}$ \\
\hline $n_{1}$ & \circled{1} & 0 & \circled{-1} & 0 \\
$n_{2}$ & 0 & 0 & 0 & 0 \\
$n_{4}$ & \circled{-1} & 0 & \circled{1} & 0 \\
$n_{5}$ & 0 & 0 & 0 & 0 \\
\end{tabular},
\end{equation*}
where $\tx{l_{1}} = n_{1}$ and $\rx{l_{1}} = n_{4}$.
It can be seen that there are four non-zero entries in $\a_{1}^{(1)} (\a_{1}^{(1)})^{T}$.
As stated in Case 1 of Lemma~\ref{lem:alfalft}, the entries on the main diagonal corresponding to $n_{1}$ and $n_{4}$ are equal to $1$.
Also, the off-diagonal entries that correspond to $n_{1}$ and $n_{4}$ are $-1$.

On the other hand, it can be verified that
\begin{equation*}
\a_{5}^{(1)} (\a_{5}^{(1)})^{T} =
\begin{tabular}{r|cccc}
 & $n_{1}$ & $n_{2}$ & $n_{4}$ & $n_{5}$ \\
\hline $n_{1}$ & 0 & 0 & 0 & 0 \\
$n_{2}$ & 0 & \circled{1} & 0 & 0 \\
$n_{4}$ & 0 & 0 & 0 & 0 \\
$n_{5}$ & 0 & 0 & 0 & 0 \\
\end{tabular},
\end{equation*}
where $\tx{l_{5}} = n_{2}$ and $\rx{l_{5}} = n_{3} = \dst{f_{1}}$.
It can be seen that there is only one non-zero entry in $\a_{5}^{(1)} (\a_{5}^{(1)})^{T}$.
As stated in Case 2 of Lemma~\ref{lem:alfalft}, the entry on the main diagonal corresponding to $n_{2}$ is equal to $1$ and all other entries are zeros.

Also, we can verify that
\begin{equation*}
\a_{1}^{(1)} (\a_{1}^{(2)})^{T} =
\begin{tabular}{c|cccc}
 & $n_{1}$ & $n_{2}$ & $n_{3}$ & $n_{4}$ \\
\hline $n_{1}$ & \circled{1} & 0 & 0 & \circled{-1} \\
$n_{2}$ & 0 & 0 & 0 & 0 \\
$n_{4}$ & \circled{-1} & 0 & 0 & \circled{1} \\
$n_{5}$ & 0 & 0 & 0 & 0
\end{tabular},
\end{equation*}
where $\tx{l_{1}} = n_{1}$ and $\rx{l_{1}} = n_{4}$, and neither $n_{1}$ nor $n_{4}$ is equal to $\dst{f_{1}}$ or $\dst{f_{2}}$.
In this case, it can be seen that there are four non-zero entries in $\a_{1}^{(1)} (\a_{1}^{(2)})^{T}$.
As stated in Case 1 of Lemma~\ref{lem:alf1alf2t}, the entry whose row and column correspond to $n_{1}$ is equal to $1$ (the same is true for $n_{4}$).
Also, the two entries whose row and column respectively correspond to $n_{1}$ and $n_{4}$ are equal to $-1$.

On the other hand, it can be verified that
\begin{equation*}
\a_{4}^{(1)} (\a_{4}^{(2)})^{T} =
\begin{tabular}{c|cccc}
 & $n_{1}$ & $n_{2}$ & $n_{3}$ & $n_{4}$ \\
\hline $n_{1}$ & 0 & 0 & 0 & 0 \\
$n_{2}$ & 0 & \circled{1} & 0 & 0 \\
$n_{4}$ & 0 & 0 & 0 & 0 \\
$n_{5}$ & 0 & \circled{-1} & 0 & 0
\end{tabular},
\end{equation*}
where $\tx{l_{4}} = n_{2}$ and $\rx{l_{4}} = n_{5} = \dst{f_{2}}$.
It can be seen that there are only two non-zero entries in $\a_{4}^{(1)} (\a_{4}^{(2)})^{T}$.
As stated in Case 2 of Lemma~\ref{lem:alf1alf2t}, the entry whose row and column correspond to $n_{2}$ is equal to $1$ .
Also, the entry whose row and column respectively correspond to $n_{5}$ and $n_{2}$ is equal to $1$.

Finally, it can be verified that
\begin{equation*}
\a_{6}^{(1)} (\a_{6}^{(2)})^{T} =
\begin{tabular}{c|cccc}
 & $n_{1}$ & $n_{2}$ & $n_{3}$ & $n_{4}$ \\
\hline $n_{1}$ & 0 & 0 & 0 & 0 \\
$n_{2}$ & 0 & 0 & 0 & 0 \\
$n_{4}$ & 0 & 0 & 0 & 0 \\
$n_{5}$ & 0 & 0 & \circled{-1} & 0
\end{tabular},
\end{equation*}
where $\tx{l_{6}} = n_{5} = \dst{f_{2}}$ and $\rx{l_{6}} = n_{3} = \dst{f_{1}}$.
Thus, as stated in Case 3 of Lemma~\ref{lem:alf1alf2t}, we have that the entry whose row and column respectively correspond to $n_{5}$ and $n_{3}$ is equal to $-1$.

\subsection{Distributed Computation of the Primal Newton Direction} \label{sec:Primal_Newton_Dir}
By solving (\ref{eqn_newton_linearsys2}), we have
\begin{align}
\label{eqn_newton_linsys2_1} & \Dyt^{k} = -\tHc_{k}^{-1} (\nabla f(\yt^{k}) + \Mt^{T} \wt^{k}), \\
\label{eqn_newton_linsys2_2} & (\Mt \tHc_{k}^{-1} \Mt^{T}) \wt^{k} = -\Mt \tHc_{k}^{-1} \nabla f(\yt^{k}).
\end{align}
Now, consider the Hessian matrix $\tHc_{k}$ of Problem R2-MRFC.
As mentioned earlier, since $f(\y^{k})$ is separable based on links, the Hessian matrix $\tHc_{k}$ for the rearranged Problem R2-MRFC has the following {\em block diagonal} structure:
\begin{equation*}
\tHc_{k} = \diag{\S, \X_{1}, \ldots, \X_{L}} \in \mathbb{R}^{(L+1)F \times (L+1)F},
\end{equation*}
where $\S$ is a diagonal matrix defined as $\S \triangleq \diag{ -tU''_{1}(s_{1})+\frac{1}{s_{1}^{2}}, \ldots, -tU''_{F}(s_{F})+\frac{1}{s_{F}^{2}} } \in \mathbb{R}^{F\times F}$; and where $\X_{l} \in \mathbb{R}^{F\times F}$ is a symmetric matrix with entries defined as follows:
\begin{equation} \label{eqn_Xl_struc}
(\X_{l})_{f_{1},f_{2}} =
\begin{cases}
\frac{1}{\delta_{l}^{2}} + \frac{1}{\big(x_{l}^{(f_{1})}\big)^{2}} & \text{if $f_{1}=f_{2}$},\\
\frac{1}{\delta_{l}^{2}} & \text{if $f_{1} \ne f_{2}$}.
\end{cases}
\end{equation}
It then follows from the block diagonal structure of $\tHc_{k}$ that
\begin{equation*}
\tHc_{k}^{-1} = \diag{\S^{-1}, \X_{1}^{-1}, \ldots, \X_{L}^{-1}}.
\end{equation*}
Noting that $\S^{-1}$ is diagonal, we have $\S^{-1} = \diag{\frac{1}{-tU''_{1}(s_{1})+\frac{1}{s_{1}^{2}}}, \ldots, \frac{1}{-tU''_{F}(s_{F})+\frac{1}{s_{F}^{2}}} }$.
Notice further that each source node $f$ has the knowledge of $s_{f}$, which implies the following result:
\begin{lem}[Distributedness in computing $\S^{-1}$] \label{lem:src_inv_dist}
The computation of $\S^{-1}$ does not require global information and can be computed source node-wise in a distributed fashion.
Moreover, the inverse of $\S$ is given by $\S^{-1} = \diag{\frac{1}{-tU''_{1}(s_{1})+\frac{1}{s_{1}^{2}}}, \ldots, \frac{1}{-tU''_{F}(s_{F})+\frac{1}{s_{F}^{2}}} }$.
\end{lem}

Next, we consider the computation of $\X_{l}^{-1}$.
Note that $\X_{l}$ only involves variables $x_{l}^{(f)}$, $f=1,\ldots,F$, which are available at each link locally.
Hence, we have the following result:
\begin{lem}[Distributedness of computing $\X_{l}^{-1}$] \label{lem:link_inv_dist}
The computation of $\X_{l}^{-1}$ does not require global information and can be computed link-wise in a distributed fashion.
\end{lem}

For convenience, we define a new vector $\xh_{l} \triangleq \Big[x_{l}^{(1)}, \ldots, x_{l}^{(F)}, \delta_{l} \Big]^{T} \in \mathbb{R}^{F+1}$.
Note that this $(F+1)$-dimensional vector only has $F$ degrees-of-freedom (DoF) and its $L_{1}$-norm (noting positive components due to the barrier function) is a constant $C_{l}$.
Then, by exploiting the special structure of $\X_{l}$ in (\ref{eqn_Xl_struc}), we can show that $\X_{l}^{-1}$ can be computed in closed-form as follows:
\begin{thm}[Closed-form expression for $\X_{l}^{-1}$] \label{thm:Xl_closedform}
The entries of $\X_{l}^{-1}$ can be computed in closed-form as follows:
\begin{equation} \label{eqn_Xl_closedform}
(\X_{l}^{-1})_{f_{1}f_{2}} =
\begin{cases}
\big(x_{l}^{(f_{1})}\big)^{2} \bigg(1-\frac{\big(x_{l}^{(f_{1})}\big)^{2}}{\|\xh_{l}\|^{2}}\bigg) & \text{if $1\leq f_{1}=f_{2} \leq F$}, \\
-\frac{\big(x_{l}^{(f_{1})}x_{l}^{(f_{2})}\big)^{2}}{\|\xh_{l}\|^{2}} & \text{if $1\leq f_{1},f_{2} \leq F$, $f_{1} \ne f_{2}$}.
\end{cases}
\end{equation}
\end{thm}
The basic idea of the proof of Theorem~\ref{thm:Xl_closedform} is based on a keen observation of the decomposable structure of $\X_{l}$ in (\ref{eqn_Xl_struc}) and the Sherman--Morrison--Woodbury formula \cite{Bazaraa_Sherali_Shetty_93:NLP}.
We relegate the details of the proof to Appendix~\ref{appdx:Xl_closedform}.

Combining Lemma~\ref{lem:src_inv_dist}, Lemma~\ref{lem:link_inv_dist}, Theorem~\ref{thm:Xl_closedform} and all related discussions earlier, we can conclude that
$\tHc_{k}^{-1}$, the inverse of the Hessian matrix of Problem R2-MRFC, can be computed distributedly as shown in the following theorem:
\begin{thm} \label{thm:primal_newton_direct}
Given dual variables $\wt$, the Newton direction $\Delta s_{f}$ and $\Delta x_{l}^{(f)}$ for each source rate $s_{f}$ and link flow rate $x_{l}^{(f)}$ can be computed using local information at each source node $s$ and link $l$, respectively.
More specifically, $\Delta s_{f}$ and $\Delta x_{l}^{(f)}$ can be computed as follows:
\begin{align}
\label{eqn_pndir1} \Delta s_{f} &= \frac{s_{f}\big(ts_{f}U'_{f}(s_{f})+1-s_{f}w_{\src{f}}^{(f)}\big)}{1-ts_{f}^{2} U''_{f}(s_{f})}, && \forall f,\\
\label{eqn_pndir2} \Delta x_{l}^{(f)} &= \big( x_{l}^{(f)} \big)^{2}
\left[ \left( 1 - \frac{(x_{l}^{(f)})^{2}}{\|\xh_{l}\|^{2}} \right)
\left( \frac{1}{x_{l}^{(f)}} - \frac{1}{\delta_{l}} + \widetilde{w}_{\tx{l}}^{(f)} - \widetilde{w}_{\rx{l}}^{(f)} \right) +
\right. \nonumber\\
& \hspace{.8in} \left. \sum_{f'=1,f'\ne f}^{F} \frac{(x_{l}^{(f')})^{2}}{\|\xh_{l}\|^{2}} \left( \frac{1}{x_{l}^{(f')}} - \frac{1}{\delta_{l}} + \widetilde{w}_{\tx{l}}^{(f')} - \widetilde{w}_{\rx{l}}^{(f')} \right) \right], && \forall l,f.
\end{align},
\end{thm}
The key steps of proving Theorem~\ref{thm:primal_newton_direct} are: (i) applying $\tHc_{k}^{-1} = \diag{\S^{-1}, \X_{1}^{-1}, \ldots, \X_{L}^{-1}}$, Lemma~\ref{lem:src_inv_dist}, Lemma~\ref{lem:link_inv_dist}, and Theorem~\ref{thm:Xl_closedform} in (\ref{eqn_newton_linsys2_1}), and (ii) exploiting the special structure of $\a_{l}^{(f)}$ and $\b^{(f)}$ to simplify the result.
We relegate the proof details to Appendix~\ref{appdx:Primal_Newton_Dir}.

\begin{rem}\label{rmk:Thm_primal_newton_dir}{\em
An important remark for Theorem~\ref{thm:primal_newton_direct} is in order.
Besides providing a closed-form expression for a distributed primal Newton direction computation, it also provides an interesting networking interpretation.
Here, we can think of the difference of the dual variables $(\widetilde{w}_{\tx{l}}^{(f)} - \widetilde{w}_{\rx{l}}^{(f)})$ in (\ref{eqn_pndir2}) as being similar to the queue length difference in the ``back-pressure'' algorithm, although $\widetilde{w}_{n}^{(f)}$ cannot be exactly interpreted as queue length (since it can be positive or negative).
Note that in (\ref{eqn_pndir2}), $1 - \frac{(x_{l}^{(f)})^{2}}{\|\xh_{l}\|^{2}}$ and $\frac{(x_{l}^{(f')})^{2}}{\|\xh_{l}\|^{2}}$ are all positive quantities.
Hence, if the positive $(\widetilde{w}_{\tx{l}}^{(f)} - \widetilde{w}_{\rx{l}}^{(f)})$-value outweigh the negative ones, i.e., the ``pressure'' on the transmitter side of link $l$ is greater than the ``pressure'' on the receiver side, then $x_{l}^{(f)}$ will be increased in the next iteration.
Note also that, unlike in first-order methods, the decision to increase or decrease $x_{l}^{(f)}$ at link $l$ considers not only the ``pressure difference'' of flow $f$ but also the ``pressure difference'' from other flows at link $l$ (via an appropriate weighting scheme as evident in (\ref{eqn_pndir2})).
}
\end{rem}

\subsection{Distributed Computation of the Dual Variables} \label{sec:Dual_Variable_Compu}
As mentioned earlier, given a primal solution $\y^{k}$ at the $k$-th iteration, the dual variables $\w^{k}$ may be computed using (\ref{eqn_newton_linsys2_2}).
However, solving for $\w^{k}$ using (\ref{eqn_newton_linsys2_2}) cannot be implemented in a distributed fashion because computing the inverse of the matrix $\Mt \tHc_{k}^{-1} \Mt^{T}$ requires global information.
In what follows, we will first study the special structure of $\Mt \tHc_{k}^{-1} \Mt^{T}$. Then, we will show that the matrix splitting scheme in \cite{Wei10:DNewton} can be generalized to compute the dual variables $\w^{k}$ as per (\ref{eqn_newton_linsys_2}).

Recall that $\Mt$ can be written in a partitioned matrix form as $\Mt = \mtrx{{cccc} \Bt & \A_{1} & \cdots & \A_{L} }$.
Hence, we can decompose $\Mt \tHc_{k}^{-1} \Mt^{T}$ as
\begin{equation} \label{eqn_MHMT_decomp}
\Mt \tHc_{k}^{-1} \Mt^{T} = \mtrx{{cccc} \Bt & \A_{1} & \cdots & \A_{L} }
\mtrx{{cccc}
\S^{-1} & & & \\
 & \X_{1}^{-1} & & \\
 & & \ddots & \\
 & & & \X_{L}^{-1}
}
\mtrx{{c} \Bt^{T} \\ \A_{1}^{T} \\ \vdots \\ \A_{L}^{T} }
= \Bt \S^{-1} \Bt^{T} + \sum_{l=1}^{L} \A_{l} \X_{l}^{-1} \A_{l}^{T}.
\end{equation}
Now, we consider each term in the decomposition in (\ref{eqn_MHMT_decomp}).
For $\Bt \S^{-1} \Bt^{T}$, since $\Bt$ and $\S^{-1}$ are diagonal, we have
\begin{equation}
\Bt \S^{-1} \Bt^{T} = \diag{\frac{1}{-tU''_{1}(s_{1})+\frac{1}{(s_{1})^{2}}} \bt^{(1)}(\bt^{(1)})^{T},\ldots,\frac{1}{-tU''_{F}(s_{F})+\frac{1}{(s_{F})^{2}}}\bt^{(F)}\bt^{(F)}},
\end{equation}
which is a block diagonal matrix.
Moreover, by Lemma~\ref{lem:btfbtft}, each bock has the following structure:
\begin{equation*}
\frac{1}{-tU''_{f}(s_{f})+\frac{1}{(s_{f})^{2}}} \diag{0 \ldots 1 \ldots 0},
\end{equation*}
where the position of the only non-zero entry $1$ corresponds to node $\src{f}$.

Next, consider the term $\sum_{l=1}^{L} \A_{l} \X_{l}^{-1} \A_{l}^{T}$, which is more involved.
From Theorem~\ref{thm:Xl_closedform}, we can decompose $\sum_{l=1}^{L} \A_{l} \X_{l}^{-1} \A_{l}^{T}$ as follows:
\begin{align} \label{eqn_AXA_decomp}
\sum_{l=1}^{L} \A_{l} \X_{l}^{-1} \A_{l}^{T} &=
\sum_{l=1}^{L} \left(
\A_{l} \mtrx{{ccc}
(x_{l}^{(1)})^{2} & & \\
 & \ddots & \\
 & & (x_{l}^{(F)})^{2}
} \A_{l}^{T}
\right) - \nonumber\\
& \sum_{l=1}^{L} \left( \frac{1}{\|\xh_{l}\|^{2}}
\A_{l} \mtrx{{cccc}
(x_{l}^{(1)})^{4} & \cdots & (x_{l}^{(1)}x_{l}^{(F)})^{2} \\
\vdots & \ddots & \vdots \\
(x_{l}^{(F)}x_{l}^{(1)})^{2} & \cdots & (x_{l}^{(F)})^{4} \\
}
\A_{l}^{T}
\right).
\end{align}
Due to the block diagonal structure, the first term in (\ref{eqn_AXA_decomp}) can be further written as
\begin{equation*}
\sum_{l=1}^{L} \left(
\A_{l} \diag{(x_{l}^{(1)})^{2},\ldots,(x_{l}^{(F)})^{2}} \A_{l}^{T}
\right) =
\sum_{l=1}^{L} \diag{
(x_{l}^{(1)})^{2} \a_{l}^{(1)} (\a_{l}^{(1)})^{T},
\ldots,
(x_{l}^{(F)})^{2} \a_{l}^{(F)} (\a_{l}^{(F)})^{T}
},
\end{equation*}
which is also a block diagonal matrix.
Thus, we can combine this term with $\Bt \S^{-1} \Bt^{T}$.
For convenience, we let $\D \triangleq \Bt \S^{-1} \Bt^{T} + \sum_{l=1}^{L} \A_{l} \diag{(x_{l}^{(1)})^{2},\ldots,(x_{l}^{(F)})^{2}} \A_{l}^{T}$.
Clearly, $\D$ is also block diagonal, and can be written as $\D = \diag{\D_{1},\ldots,\D_{F}}$.
Then, by using Lemma~\ref{lem:alfalft}, we obtain the following result, where the proof is relegated to Appendix~\ref{appdx:D_f}.
\begin{lem} \label{lem:D_f}
The matrix $\D$ is block diagonal and each block $\D_{f}$ on the main diagonal has the following structure:
\begin{itemize}
\item The diagonal entries $(\D_{f})_{ii}$ are given by
\begin{equation*}
(\D_{f})_{ii} =
\begin{cases}
\sum_{l\in \Out{n} \cup \In{n}} (x_{l}^{(f)})^{2} + \frac{1}{-tU''_{f}(s_{f})+\frac{1}{(s_{f})^{2}}} & \text{if row $i$ corresponds to node $n$ and $n=\src{f}$}, \\
\sum_{l\in \Out{n} \cup \In{n}} (x_{l}^{(f)})^{2} & \text{otherwise}.
\end{cases}
\end{equation*}

\item The off-diagonal entries of $(\D_{f})_{ij}$, $i\ne j$, are given by
\begin{equation*}
(\D_{f})_{ij} =
\begin{cases}
-\sum_{l \in \Gamma(n_{1},n_{2})}(x_{l}^{(f)})^{2} & \text{if row $i$ and column $j$ correspond to two connected nodes $n_{1}$ and $n_{2}$}, \\
0 & \text{otherwise},
\end{cases}
\end{equation*}
where $\Gamma(n_{1},n_{2}) \triangleq \{l \in \mathcal{L}: \tx{l} = n_{1} \text{ and } \rx{l} = n_{2}, \text{ or } \tx{l} = n_{2} \text{ and } \rx{l} = n_{1} \}$.
\end{itemize}
\end{lem}

Next, we study the second term in (\ref{eqn_AXA_decomp}), denoted as $\C$,
which is symmetric and can be shown to have the following partitioned structure using Theorem~\ref{thm:Xl_closedform}:
\begin{align*}
\C &\triangleq \sum_{l=1}^{L} \left( \frac{1}{\|\xh_{l}\|^{2}}
\A_{l} \mtrx{{cccc}
(x_{l}^{(1)})^{4} & \cdots & (x_{l}^{(1)}x_{l}^{(F)})^{2} \\
\vdots & \ddots & \vdots \\
(x_{l}^{(F)}x_{l}^{(1)})^{2} & \cdots & (x_{l}^{(F)})^{4} \\
}
\A_{l}^{T}
\right) \\
 &=
\mtrx{{cccc}
\Dh_{1} & \G_{12} & \cdots & \G_{1F} \\
\G_{21} & \Dh_{2} & \cdots & \G_{2F} \\
\vdots & \vdots & \ddots & \vdots \\
\G_{F1} & \G_{F2} & \cdots & \Dh_{F}
},
\end{align*}
where
\begin{align*}
& \Dh_{f} = \sum_{l=1}^{L} \frac{(x_{l}^{(f)})^{4}}{\|\xh_{l}\|^{2}} \a_{l}^{(f)} (\a_{l}^{(f)})^{T}, && f=1,\ldots,F, \\
& \G_{f_{1}f_{2}} = \sum_{l=1}^{L} \frac{(x_{l}^{(f_{1})}x_{l}^{(f_{2})})^{2}}{\|\xh_{l}\|^{2}} \a_{l}^{(f_{1})} (\a_{l}^{(f_{2})})^{T}, && f_{1},f_{2}=1,\ldots,F, f_{1} \ne f_{2}.
\end{align*}
Noting the similarity between $\Dh_{f}$ and $\D_{f}$, and by using Lemma~\ref{lem:alfalft} and following a similar derivation to that for Lemma~\ref{lem:D_f}, we obtain the following result for characterizing $\Dh_{f}$, where we omit the proof to avoid repetition.
\begin{lem} \label{lem:Dh_f}
The matrix $\Dh_{f}$ has the following structure:
\begin{itemize}
\item The diagonal entries $(\Dh_{f})_{ii}$ are given by
\begin{equation*}
(\Dh_{f})_{ii} =
\sum_{l\in \Out{n} \cup \In{n}} \frac{(x_{l}^{(f)})^{4}}{\|\xh_{l}\|^{2}}.
\end{equation*}

\item The off-diagonal entries of $(\Dh_{f})_{ij}$, $i\ne j$, are given by
\begin{equation*}
(\Dh_{f})_{ij} =
\begin{cases}
-\sum_{l\in \Gamma(n_{1},n_{2})}\frac{(x_{l}^{(f)})^{4}}{\|\xh_{l}\|^{2}} & \text{if row $i$ and column $j$ correspond to two connected nodes $n_{1},n_{2}$}, \\
0 & \text{otherwise}.
\end{cases}
\end{equation*}
\end{itemize}
\end{lem}
Using Lemma~\ref{lem:alf1alf2t}, we can also characterize the structure of $\G_{f_{1}f_{2}}$ as stated in Lemma~\ref{lem:G_f1f2} below, where the proof follows that of Lemma~\ref{lem:D_f} and is therefore omitted for the sake of brevity.
\begin{lem} \label{lem:G_f1f2}
The matrix $\G_{f_{1}f_{2}}$ has the following structure:
\begin{equation*}
(\G_{f_{1}f_{2}})_{ij} =
\begin{cases}
\sum_{l\in \Out{n}\cup\In{n}} \frac{(x_{l}^{(f_{1})}x_{l}^{(f_{2})})^{2}}{\|\xh_{l}\|^{2}} & \text{if row $i$ and column $j$ correspond to the same node $n$}, \\
-\sum_{l\in\Gamma(n_{1},n_{2})}\frac{(x_{l}^{(f_{1})}x_{l}^{(f_{2})})^{2}}{\|\xh_{l}\|^{2}} & \text{if row $i$ and column $j$ correspond to two connected nodes $n_{1},n_{2}$}, \\
0 & \text{otherwise}.
\end{cases}
\end{equation*}
\end{lem}
So far, we have characterized the structures of $\Dh_{f}$ and $\G_{f_{1}f_{2}}$.
Hence, the structure of $\C$ is also known.
Finally, recall that $\Mt \tHc_{k}^{-1} \Mt^{T} = \D - \C$.
Therefore, combining the previous derivations, we have the following result for the structure property of $\Mt \tHc_{k}^{-1} \Mt^{T}$.
\begin{thm} \label{thm:MHMT}
The matrix $\Mt \tHc_{k}^{-1} \Mt^{T}$ can be written as the following partitioned matrix:
\begin{equation*}
\Mt \tHc_{k}^{-1} \Mt^{T} =
\mtrx{{llll}
\D_{1}-\Dh_{1} & -\G_{12} & \cdots & -\G_{1F} \\
-\G_{21} & \D_{2}-\Dh_{2} & \cdots & -\G_{2F} \\
\vdots & \vdots & \ddots & \vdots \\
-\G_{F1} & -\G_{F2} & \cdots & \D_{F}-\Dh_{F}
},
\end{equation*}
where the structures properties of the matrices $\D_{f}$, $\Dh_{f}$, and $\G_{f_{1}f_{2}}$ are specified in Lemma~\ref{lem:D_f}, Lemma~\ref{lem:Dh_f}, and Lemma~\ref{lem:G_f1f2}, respectively.
\end{thm}

Armed with Theorem~\ref{thm:MHMT}, we are now in a position to design a distributed iterative scheme to compute the dual variables $\wt_{k}$ using matrix splitting techniques.
To this end, we first introduce some preliminary results in matrix splitting theory.

Historically, the idea of matrix splitting has its origin in designing iterative schemes to solve linear equation systems \cite{Woznicki01:Mtrx_Split}.
Consider a consistent linear equation system $\F \z = \d$, where $\F \in \mathbb{R}^{n\times n}$ is a nonsingular matrix and $\z,\d \in \mathbb{R}^{n}$.
Now, suppose that $\F$ is split into a nonsingular matrix $\F_{1}$ and another matrix $\F_{2}$ according to $\F = \F_{1}-\F_{2}$.
Also, let $\z^{0}$ be an arbitrary starting vector.
Then, a sequence of approximate solutions can be generated by using the following iterative scheme:
\begin{equation} \label{eqn_MtrxSplit_Appr_Soln}
\z^{k+1} = (\F_{1}^{-1} \F_{2}) \z^{k} + \F_{1}^{-1}\d, \quad k\geq 0.
\end{equation}
Generally, $\F_{1}$ should be an easily invertible matrix (e.g., diagonal, etc).
It can be shown that this iterative method is convergent to the unique solution $\z = \F^{-1}\b$ if and only if the spectral radius of the matrix $\F_{1}^{-1}\F_{2}$ is less than one, i.e., $\rho(\F_{1}^{-1}\F_{2}) < 1$, where $\rho(\cdot)$ represents the spectral radius of a matrix.
The following result provides a sufficient condition for $\rho(\F_{1}^{-1}\F_{2}) < 1$ (see \cite{Woznicki01:Mtrx_Split,Wei10:DNewton} for more details).
\begin{lem} \label{lem:Mtrx_Split_Conv_Suff_Cond}
Suppose that $\F$ is a real symmetric matrix. If both matrices $\F_{1}+\F_{2}$ and $\F_{1}-\F_{2}$ are positive definite, then $\rho(\F_{1}^{-1}\F_{2}) < 1$.
\end{lem}
Lemma~\ref{lem:Mtrx_Split_Conv_Suff_Cond} suggests that the convergence property of a given matrix splitting scheme can be verified by checking for the positive definiteness of the identified matrix.
The following result provides a sufficient condition for checking positive definiteness based on diagonal dominance \cite[Corollary 7.2.3]{Horn_Johnson90:Mtrx_Thry}:
\begin{lem} \label{lem:PD}
If a symmetric matrix $\Q$ is strictly diagonally dominant, i.e., $|(\Q)_{ii}| > \sum_{j \ne i} |(\Q)_{ij}|$, and if $(\Q)_{ii} > 0$ for all $i$, then $\Q$ is positive definite.
\end{lem}
We are now ready to use the matrix splitting scheme in (\ref{eqn_MtrxSplit_Appr_Soln}) to compute $\wt^{k}$.
First, we let $\mLambda$ be the diagonal matrix having the same main diagonal of $\Mt \tHc_{k}^{-1} \Mt^{T}$, i.e.,
\begin{equation} \label{eqn_Lambda}
\mLambda_{k} = \diag{\diagv{\Mt \tHc_{k}^{-1} \Mt^{T}}}.
\end{equation}
We let $\mOmega$ denote the matrix containing the remaining entries after subtracting $\mLambda_{k}$ from $\Mt \tHc_{k}^{-1} \Mt^{T}$, i.e.,
\begin{equation} \label{eqn_Omega}
\mOmega_{k} = \Mt \tHc_{k}^{-1} \Mt^{T} - \mLambda_{k}.
\end{equation}
Further, we define a diagonal matrix $\mOmegab_{k}$ where the diagonal entries are defined by
\begin{equation} \label{eqn_Omegab}
(\mOmegab_{k})_{ii} = \sum_{j} |(\mOmega_{k})_{ij}|.
\end{equation}
Then, we can split $\Mt \tHc_{k}^{-1} \Mt^{T}$ as $(\mLambda_{k} + \alpha \mOmegab_{k})-(\alpha \mOmegab_{k} - \mOmega_{k})$, where $\alpha > \frac{1}{2}$ is a parameter that serves the purpose of tuning convergence performance.
Based on this splitting scheme, we have the following result:
\begin{thm} \label{thm_Distr_Dual_Update}
Consider the matrix splitting scheme $\Mt \tHc_{k}^{-1} \Mt^{T}$ as $\Mt \tHc_{k}^{-1} \Mt^{T} = (\mLambda_{k} + \alpha \mOmegab_{k}) - (\alpha \mOmegab_{k} - \mOmega_{k})$, where $\mLambda_{k}$, $\mOmega_{k}$, and $\mOmegab_{k}$ are defined in (\ref{eqn_Lambda}), (\ref{eqn_Omega}), and (\ref{eqn_Omegab}), respectively.
Then, the following sequence $\{\wt^{k}\}$ generated by
\begin{equation} \label{eqn_Distr_Dual_Update}
\wt^{k+1} = (\mLambda_{k} + \alpha \mOmegab_{k})^{-1} (\alpha \mOmegab_{k} - \mOmega_{k}) \wt^{k} + (\mLambda_{k} + \alpha \mOmegab_{k})^{-1} (-\Mt\tHc_{k}^{-1}\nabla f(\yt^{k}))
\end{equation}
converges to the solution of (\ref{eqn_newton_linsys2_2}) as $k\rightarrow \infty$.
\end{thm}
By Lemmas~\ref{lem:Mtrx_Split_Conv_Suff_Cond} and~\ref{lem:PD}, the key to proving Theorem~\ref{thm_Distr_Dual_Update} is to verify that both the sum and difference of the two components in the splitting scheme are strictly diagonally dominant.
We relegate the proof details to Appendix~\ref{appdx:Distr_Dual_Update}.

\begin{rem}{\em
The matrix splitting scheme in Theorem~\ref{thm_Distr_Dual_Update} is inspired by, and is a generalization of, the scheme in \cite{Wei10:DNewton}.
The goal of both splitting schemes is to construct a diagonal nonsingular matrix ($\mLambda_{k} + \alpha \mOmegab_{k}$ in our paper) for which the inverse can be separated and easily computed by each node (as in our case) or each link (as in \cite{Wei10:DNewton}).
However, our matrix splitting scheme differs from that in \cite{Wei10:DNewton} in the following aspects.
First, since $(\mLambda_{k} + \alpha \mOmegab_{k})$ is not element-wise non-negative (c.f. \cite{Wei10:DNewton}), the definition of the matrix $\mOmegab_{k}$ in this work is different from that in \cite{Wei10:DNewton}, which also leads to a different proof.
Second, we parameterize the splitting scheme (using $\alpha$) to allow for tuning the convergence speed in (\ref{eqn_Distr_Dual_Update}), where the scheme in \cite{Wei10:DNewton} is a special case of our scheme when $\alpha=1$.
}
\end{rem}

Some comments on the parameter $\alpha$ are addressed at this point.
From (\ref{eqn_MtrxSplit_Appr_Soln}), it can be seen that the solution error decreases in magnitude approximately by a factor of $\rho(\F_{1}^{-1}\F_{2})$.
Thus, the smaller $\rho(\F_{1}^{-1}\F_{2})$ is, the faster the convergence we might expect of the iterative scheme.
To this end, we have the following result for the selection of the parameter $\alpha$.
\begin{prop} \label{prop:Alpha_Choice}
Consider two alternative matrix splitting schemes with parameters $\alpha_{1}$ and $\alpha_{2}$, respectively, satisfying $\frac{1}{2} < \alpha_{1} \leq \alpha_{2}$.
Let $\rho_{\alpha_{1}}$ and $\rho_{\alpha_{2}}$ be their spectral radii, respectively.
Then, $\rho_{\alpha_{1}} \leq \rho_{\alpha_{2}}$.
\end{prop}

Proposition~\ref{prop:Alpha_Choice} implies that in order to make the matrix splitting scheme converge faster, we should choose a smaller $\alpha$, i.e., we can let $\alpha = \frac{1}{2}+\epsilon$, where $\epsilon > 0$ is small.
The proof of Proposition~\ref{prop:Alpha_Choice} makes use of the comparison theorem in \cite{Woznicki01:Mtrx_Split} and we relegate its details to Appendix~\ref{appdx:Alpha_Choice}.

Next, we show that the matrix splitting scheme in Theorem~\ref{thm_Distr_Dual_Update} can indeed be implemented in a distributed fashion to solve the MRFC problem.
For convenience, we define two types of link sets as follows:
\begin{align*}
& \Phi(n) \triangleq \In{n} \cup \Out{n},
& \Psi(n,f) \triangleq \left\{l \in \In{n}\cup\Out{n}: \tx{l}=\dst{f} \text{ or } \rx{l}=\dst{f}\right\}.
\end{align*}
We let $\ind_{S}(a)$ denote the set indicator function, which takes value 1 if $a\in S$ and 0 otherwise.
Then, we have the following result:
\begin{thm} \label{thm:Distr_Dual_Update_Exp}
Given a primal solution $\yt^{k}$, the update of the dual variable $w_{n}^{(f)}$ can be computed using local information at each node.
More specifically, $w_{n}^{(f)}$ can be written as
\begin{equation} \label{eqn_Distr_Dual_Update_Exp}
w_{n}^{(f)}(k+1) = \frac{1}{U_{n}^{f}(k)}(V_{n,1}^{(f)}(k) + V_{n,2}^{(f)}(k) - W_{n}^{f}(k)),
\end{equation}
where $U_{n}^{(f)}(k)$, $V_{n}^{(f)}(k)$, and $W_{n}^{(f)}(k)$ are, respectively, defined as
\begin{equation} \label{eqn_Unfk}
U_{n}^{(f)}(k) \triangleq
\begin{cases}
\sum_{l \in \Phi(n)}[1+\alpha(1-\ind_{\Psi(n,f)}(l))] (x_{l}^{(f)})^{2}\Big(1-\frac{(x_{l}^{(f)})^{2}}{\|\xh_{l}\|^{2}} \Big) + & \\
\sum_{f'=1,\ne f}^{F}\Big( \sum_{l \in \Psi(n,f')} (1+\ind_{\Psi(n,f')}(l))\frac{\alpha(x_{l}^{(f)}x_{l}^{(f')})^{2}}{\|\xh_{l}\|^{2}} \Big) & \text{if $n\ne \src{f}$}, \\
\sum_{l \in \Phi(n)}[1+\alpha(1-\ind_{\Psi(n,f)}(l))] (x_{l}^{(f)})^{2}\Big(1-\frac{(x_{l}^{(f)})^{2}}{\|\xh_{l}\|^{2}} \Big) + & \\
\sum_{f'=1,\ne f}^{F}\Big( \sum_{l \in \Psi(n,f')} (1+\ind_{\Psi(n,f')}(l))\frac{\alpha(x_{l}^{(f)}x_{l}^{(f')})^{2}}{\|\xh_{l}\|^{2}} \Big) + \frac{1}{-tU''_{f}(s_{f})+\frac{1}{(s_{f})^{2}}}& \text{if $n=\src{f}$},
\end{cases}
\end{equation}

\begin{align} \label{eqn_Vn1fk}
V_{n,1}^{(f)}(k) & \triangleq
\sum_{l\in \In{n}\backslash \Psi(n,f)} (x_{l}^{(f)})^{2} \Big(1-\frac{((x_{l}^{(f)})^{2})}{\|\xh_{l}\|^{2}}\Big)(\wts_{\tx{l}}^{(f)} + \alpha\wts_{\rx{l}}^{(f)}) + \nonumber\\
& \hspace{.2in} \sum_{l\in \Out{n}\backslash \Psi(n,f)} (x_{l}^{(f)})^{2} \Big(1-\frac{((x_{l}^{(f)})^{2})}{\|\xh_{l}\|^{2}}\Big)(\wts_{\rx{l}}^{(f)} + \alpha\wts_{\tx{l}}^{(f)}) - \nonumber\\
& \hspace{.2in} \sum_{f'=1,\ne f}^{F} \Big( \sum_{l \in \Phi(n)} (1+\ind_{\Psi(n,f')}(l))\frac{\alpha(x_{l}^{(f)}x_{l}^{(f')})^{2}}{\| \xh_{l} \|^{2}} \Big)\wts_{n}^{f},
\end{align}

\begin{equation} \label{eqn_Vn2fk}
V_{n,2}^{(f)}(k) \triangleq
\sum_{f'=1,\ne f}^{F} \Big(
\Big(\sum_{l \in \In{n}} \frac{(x_{l}^{(f)}x_{l}^{(f')})^{2}}{\|\xh_{l}\|^{2}} - \sum_{l \in \Out{n}} \frac{(x_{l}^{(f)}x_{l}^{(f')})^{2}}{\|\xh_{l}\|^{2}}
 \Big)
(\wts_{\rx{l}}^{(f')} - \wts_{\tx{l}}^{(f')})
\Big),
\end{equation}

\begin{equation} \label{eqn_Wnfk}
W_{n}^{(f)}(k) \triangleq
\begin{cases}
\Big(1-\frac{x_{l}^{(f)}}{\delta_{l}}\Big)
\Big[
\sum_{l\in \Out{n}} \Big( 1-\sum_{f'=1}^{F} \frac{(x_{l}^{(f)})^{2}}{\|\xh_{l}\|^{2}} x_{l}^{(f')} \Big) - & \\
\hspace{.3in} \sum_{l\in \In{n}} \Big( 1-\sum_{f'=1}^{F} \frac{(x_{l}^{(f)})^{2}}{\|\xh_{l}\|^{2}} x_{l}^{(f')} \Big)
\Big] & \text{if $n \ne \src{f}$}, \\
\sum_{l\in \Out{n}} \Big( 1-\sum_{f'=1}^{F} \frac{(x_{l}^{(f)})^{2}}{\|\xh_{l}\|^{2}} x_{l}^{(f')} \Big) - & \\
\hspace{.3in} \sum_{l\in \In{n}} \Big( 1-\sum_{f'=1}^{F} \frac{(x_{l}^{(f)})^{2}}{\|\xh_{l}\|^{2}} x_{l}^{(f')} \Big)
\Big] + \frac{s_{f}(1+ts_{f}U'_{f}(s_{f}))}{ts_{f}^{2}U''_{f}(s_{f})-1} & \text{if $n = \src{f}$}. \\
\end{cases}
\end{equation}
\end{thm}
Theorem~\ref{thm:Distr_Dual_Update_Exp} can be proved by computing the element-wise expansion of (\ref{eqn_Distr_Dual_Update}).
We relegate the proof details to Appendix~\ref{appdx:Distr_Dual_Update_Exp}.

\begin{rem} {\em
There are several interesting remarks pertaining to Theorem~\ref{thm:Distr_Dual_Update_Exp}.
First, it can be seen from (\ref{eqn_Unfk}), (\ref{eqn_Vn1fk}), (\ref{eqn_Vn2fk}), and (\ref{eqn_Wnfk}) that all the information needed to update $w_{n}^{(f)}$ are either locally available at node $n$ or at links that touch node $n$.
This confirms that the matrix splitting scheme can be distributedly implemented.
Second, suppose that $\alpha=1$; then it can be verified that $V_{n,1}^{(f)}$ involves a difference of quadratic terms of flow $x_{l}^{(f)}$ coming into and going out of node $n$.
This bears some resemblance to the dual update scheme in the subgradient method (cf. (\ref{eqn_sg_dual_update})).
The key difference is that all the quantities here are of second-order and are weighted by $\wts_{tx{l}}^{(f)} - \wts_{\rx{l}}^{(f)}$, which can be loosely interpreted as ``back-pressure'' (see Remark~\ref{rmk:Thm_primal_newton_dir}).
Likewise, $V_{n,2}^{(f)}$ also involves a similar ``back-pressure'' weighting mechanism.
But unlike $V_{n,1}^{(f)}$, the second-order quantities in $V_{n,2}^{(f)}$ are related to cross-session flow products $x_{l}^{(f)}x_{l}^{(f')}$.
Third, although the dual update scheme within a second-order method is more complex at each node, the more rapid convergence rate of a second-order method, with its accompanying less information exchange, outweigh this local computational cost increase.
}
\end{rem}

\subsection{Implementation of the Distributed Newton Method} \label{sec:DNewton_implement}
Although we have derived the main elements of a distributed computational scheme for obtaining the primal Newton direction and for updating the dual variables, which are key parts in our proposed distributed Newton method, there are a few open questions yet to be answered for practical implementations.
In what follows, we will discuss these issues, namely, the scale of information exchange, stopping criterion, step-size selection, etc.

\subsubsection{Information Exchange Scale Analysis}
We now analyze the required information exchange in our proposed distributed Newton method.
We first consider the primal Newton direction update.
From Theorem~\ref{thm:primal_newton_direct}, we can see that to compute $\Delta s_{f}$, we need $s_{f}$ and $\wts_{\src{f}}^{(f)}$.
Since $s_{f}$ is available at $\src{f}$, we can see from Theorem~\ref{thm:Distr_Dual_Update_Exp} that $\wts_{\src{f}}^{(f)}$ can also be computed at $\src{f}$.
Hence, there is {\em no} need for any information exchange in computing $\Delta s_{f}$.

\begin{figure}[t!]
    \begin{minipage}[t]{0.45\linewidth}
        \centering
        \includegraphics[width=2.5in]{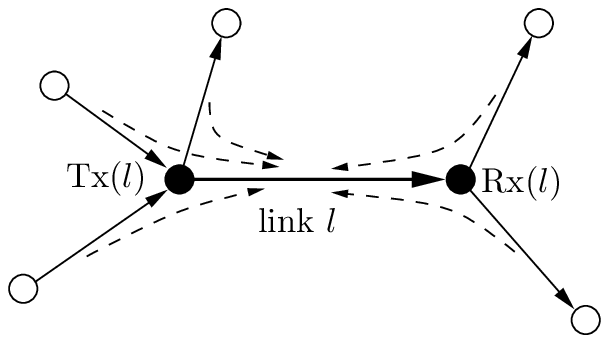}
        \caption{Information exchange for computing $\Delta x_{l}^{(f)}$, which only requires exchanging information from links one-hop away from link $l$.} \label{fig_info_exch_link}
    \end{minipage}%
    \hspace{.1in}
    \begin{minipage}[t]{0.45\linewidth}
        \centering
        \includegraphics[width=2.5in]{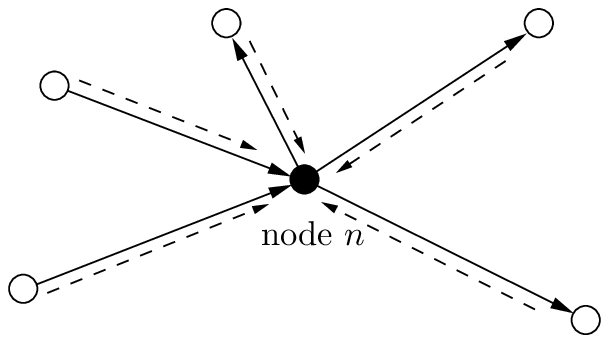}
        \caption{Information exchange for computing $\wts_{n}^{(f)}$, which only requires exchanging information from nodes one-hop away from node $n$.} \label{fig_info_exch_node}
    \end{minipage}
\end{figure}

To compute $\Delta x_{l}^{(f)}$, we can see from (\ref{eqn_pndir2}) that we need $x_{l}^{f'}$, $f'=1,\ldots,F$, $\wts_{\tx{l}}^{(f')}$, and $\wts_{\rx{l}}^{(f')}$.
Clearly, $x_{l}^{f'}$, $f'=1,\ldots,F$ are already available at link $l$.
From Theorem~\ref{thm:Distr_Dual_Update_Exp}, we can see that $\wts_{\tx{l}}^{(f')}$ and $\wts_{\rx{l}}^{(f')}$ can also be computed using flow and dual information with respect to links that share $\tx{l}$ and $\rx{l}$.
This implies that computing $\Delta x_{l}^{(f)}$ only requires exchanging information {\em one-hop} away from link $l$, as shown in Fig.~\ref{fig_info_exch_link}.

Next, consider the updating of dual variables.
From Theorem~\ref{thm:Distr_Dual_Update_Exp}, we can see that to compute $\wts_{n}^{(f)}$, we need $x_{l}^{(f')}$, $\wts_{\tx{l}}^{(f')}$, and $\wts_{\rx{l}}^{(f')}$, where $l \in \Phi(n)$, $f'=1,\ldots,F$.
It is clear that $x_{l}^{(f')}$, $l\in \Phi(n)$, is readily available at node $n$.
On the other hand, $\wts_{\tx{l}}^{(f')}$ and $\wts_{\rx{l}}^{(f')}$ are either available at node $n$ itself or are available at nodes one-hop away from node $n$.
This implies that computing $\wts_{n}^{(f)}$ only requires exchanging information from nodes {\em one-hop} away from node $n$, as shown in Fig.~\ref{fig_info_exch_node}.

Two interesting remarks are in order.
First, although the MRFC problem is more complex than the pure flow control problem in \cite{Wei10:DNewton}, the information exchange required for the distributed Newton algorithm for MRFC turns out to be more decentralized than that in \cite{Wei10:DNewton}.
More specifically, the information exchange for MRFC is from entities at most one-hop away, while in the pure flow control problem in \cite{Wei10:DNewton}, each source node needs to send information to all the links on its predefined route.
This somewhat surprising result can be loosely explained by the fact that by allowing multi-path routing, the routing decision is automatically determined by the node ``pressure'' as described in Remark~\ref{rmk:Thm_primal_newton_dir} at each node, thus alleviating the burden of exchanging information along the fixed routes.
Second, we can see that our distributed Newton method requires a similar scale of information exchange to that in the subgradient method.

\subsubsection{Initialization of the Algorithm}
Another open question in the implementation is how to initialize the algorithm.
One simple solution is as follows.
Each $\src{f}$ can choose an initial value $\epsilon_{f}$ and equally distribute $\epsilon_{f}$ along all of its outgoing links.
Also, if each intermediate node has multiple outgoing links, then the sum of its incoming traffic will be equally distributed along each outgoing link as well.
Clearly, if $\epsilon_{f}$, $\forall f$, are small enough, then the constraint $\sum_{f=1}^{F}x_{l}^{(f)}\leq C_{l}$ can be satisfied.
The initial values of dual variables can be chosen arbitrarily since they are unrestricted (e.g., a simple choice is to set $\wts_{n}^{(f)} = 1$ if $n\ne \dst{f}$ and $\wts_{n}^{(f)}=0$ if $n=\dst{f}$).

\subsubsection{Stopping Criterion}
Since the Newton method enjoys a quadratic rate of convergence, a simple stopping rule is to let all sources and links run the algorithm for a fixed amount of time.
If the time duration is long enough for a given maximum sized network, then due to the rapid convergence speed, by the time the clock expires, it is with high probability that the algorithm will have converged to a very near-optimal solution.

Another more sophisticated way to stop the algorithm can be based on the so-called Newton decrement \cite{Boyd_Vandenberghe04:Cnvx_Opt}.
In Newton methods, at a given primal vector $\yt^{k}$, the Newton decrement is defined as \cite{Boyd_Vandenberghe04:Cnvx_Opt}
\begin{equation}
\lambda(\yt^{k}) = \sqrt{(\Delta \yt^{k})^{T} \tHc_{k}\Delta \yt^{k}},
\end{equation}
which measures the decrease in the objective function value at each iteration.
Thus, we can use $\lambda(\yt^{k}) \leq \epsilon$ as a stopping criterion, where $\epsilon$ is a predefined error tolerance.
The following result shows that $\lambda(\yt^{k})$ can also be computed in a distributed fashion.
Again, for ease of notation, we omit the iteration index $k$.
\begin{prop} \label{prop:Distr_Newton_Decre}
The Newton decrement $\lambda(\yt)$ can be computed as
\begin{equation} \label{eqn_Distr_Newton_Decre}
\lambda(\yt) = \bigg(
\sum_{f=1}^{F}(\Delta s_{f})^{2} \Big( -tU''_{f}(s_{f})+\frac{1}{(s_{f})^{2}} \Big)
+\sum_{l=1}^{L} \bigg[
\sum_{f=1}^{F} \Big( \frac{\Delta x_{l}^{(f)}}{x_{l}^{(f)}} \Big)^{2}
+\frac{1}{\delta_{l}} \Big( \sum_{f=1}^{F} \Delta x_{l}^{(f)} \Big)^{2}
\bigg]
\bigg)^{\frac{1}{2}}.
\end{equation}
\end{prop}
We remark that since (\ref{eqn_Distr_Newton_Decre}) is separable with respect to each source node and each link, each source can compute the quantity $(\Delta s_{f})^{2} \Big( -tU''_{f}(s_{f})+\frac{1}{(s_{f})^{2}} \Big)$ and each link can compute the quantity $\Big( \frac{\Delta x_{l}^{(f)}}{x_{l}^{(f)}} \Big)^{2}
+\frac{1}{\delta_{l}} \Big( \sum_{f=1}^{F} \Delta x_{l}^{(f)} \Big)^{2}$.
Therefore, $\lambda(\yt^{k})$ can be computed distributedly using only local information.
The proof of Proposition~\ref{prop:Distr_Newton_Decre} is based on the decomposition structure of $\tHc_{k}$ and we relegate the proof details to Appendix~\ref{appdx:Distr_Newton_Decre}.

To compute the Newton decrement, we can see from (\ref{eqn_Distr_Newton_Decre}) that each source needs $s_{f}$ and $\Delta s_{f}$ and each link needs $x_{l}^{(f)}$ and $\Delta x_{l}^{(f)}$.
From earlier discussions on $\Delta s_{f}$, we can conclude that no information exchange is required at each source node.
Also, from earlier discussions on $\Delta x_{l}^{(f)}$, we know that at most one-hop information exchange is required in this regard.
However, to allow every source and link to compute the final value of the Newton decrement, every source and link will need to broadcast a packet containing the value of $(\Delta s_{f})^{2} \Big( -tU''_{f}(s_{f})+\frac{1}{(s_{f})^{2}} \Big)$ and $\Big( \frac{\Delta x_{l}^{(f)}}{x_{l}^{(f)}} \Big)^{2}
+\frac{1}{\delta_{l}} \Big( \sum_{f=1}^{F} \Delta x_{l}^{(f)} \Big)^{2}$, respectively, to the network.
Thus, we can see that the more accurate termination time is obtained at the expense of a larger scale of information exchange across the network.

\subsubsection{Step-Size Selection}
As in the classical Newton method \cite{Bazaraa_Sherali_Shetty_93:NLP,Boyd_Vandenberghe04:Cnvx_Opt}, when the iterates $\{\yt^{k}\}$ approach a close neighborhood of an optimal solution, using a fixed step size $\pi^{k}=1$ gives us the so-called {\em quadratic rate of convergence}, which is very efficient.
If the iterates $\{\yt^{k}\}$ is far from an optimal solution (which is also called ``damped Newton phase'' and can be measured by $\|\nabla f(\y^{k})\|_{2}$ -- see \cite{Boyd_Vandenberghe04:Cnvx_Opt} for more details), then some inexact line search methods, such as the ``Armijo rule'' \cite{Bazaraa_Sherali_Shetty_93:NLP} (also called ``backtracking line search'' in \cite{Boyd_Vandenberghe04:Cnvx_Opt}) or the step-size rule in \cite{Wei10:DNewton} can be used.
Due to the inexactness of these line search methods, the theoretical convergence rate would be sub-quadratic, but still in theory and practice, superlinear, and so, much faster than subgradient-type methods.

To conclude this section, we summarize our distributed Newton method for the MRFC problem in Algorithm~\ref{alg_DNewton}.
\begin{algorithm}[t!] {
\caption{Distributed Newton Method for Solving MRFC} \label{alg_DNewton}
\begin{algorithmic}[1]
\algsetup{
linenosize=\footnotesize,
linenodelimiter=.
}
\REQUIRE
\STATE Each source and link: Choose some appropriate values of $s_{f}$ and $x_{l}^{(f)}$, $\forall f$.
\STATE Each node: Choose appropriate values of dual variables $\wts_{n}^{(f)}$, $\forall f$.
\ENSURE
\STATE Update the primal Newton directions $\Delta s_{f}$ and $\Delta x_{l}^{(f)}$ using (\ref{eqn_pndir1}) and (\ref{eqn_pndir2}) at each source node and link, respectively.
\STATE Update the dual variables $\wts_{n}^{(f)}$ using (\ref{eqn_Distr_Dual_Update_Exp}) at each node.
\STATE Terminate the algorithm if some predefined running-time limit is reached or if the Newton decrement criterion is satisfied. Otherwise, go to Step 3.
\end{algorithmic} }
\end{algorithm}

\section{Numerical Results} \label{sec:numerical}
\begin{figure}[t!]
\centering
\includegraphics[width=3.2in]{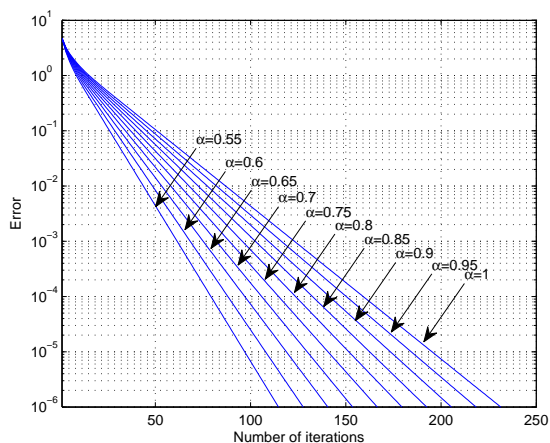}
\caption{The error between the true solution of $\w^{k}$ in Eq. (\ref{eqn_newton_linsys2_2}) and our matrix-splitting based iterative computation scheme.} \label{fig_MatrixSplitting53}
\end{figure}

In this section, we present some pertinent numerical results for our proposed distributed Newton method.
First, we examine the convergence speed of the parameterized matrix splitting scheme in Section~\ref{sec:Dual_Variable_Compu}.
We use a 10-node 3-session network as an example.
The initial values of $\w^{k}$ is set to all ones.
We vary $\alpha$ from 0.55 to 1.
The iterative scheme is stopped when the error between the true solution of $\w^{k}$ in Eq. (\ref{eqn_newton_linsys2_2}) and our matrix-splitting based iterative computation scheme is less than $1\times 10^{6}$.
The error is shown in Fig.~\ref{fig_MatrixSplitting53} (in log scale).
We can see that for all values of $\alpha$, the error decreases exponentially fast.
Also, the smaller the value of $\alpha$, the faster the convergence speed.
More specifically, when $\alpha=0.55$, the number of iterations is approximately half of that when $\alpha=1$ (115 vs. 232).
This confirms our theoretical analysis in Proposition~\ref{prop:Alpha_Choice}.

\begin{figure*}[t!]
    \begin{minipage}[t]{0.32\linewidth}
        \centering
        \includegraphics[width=2.2in]{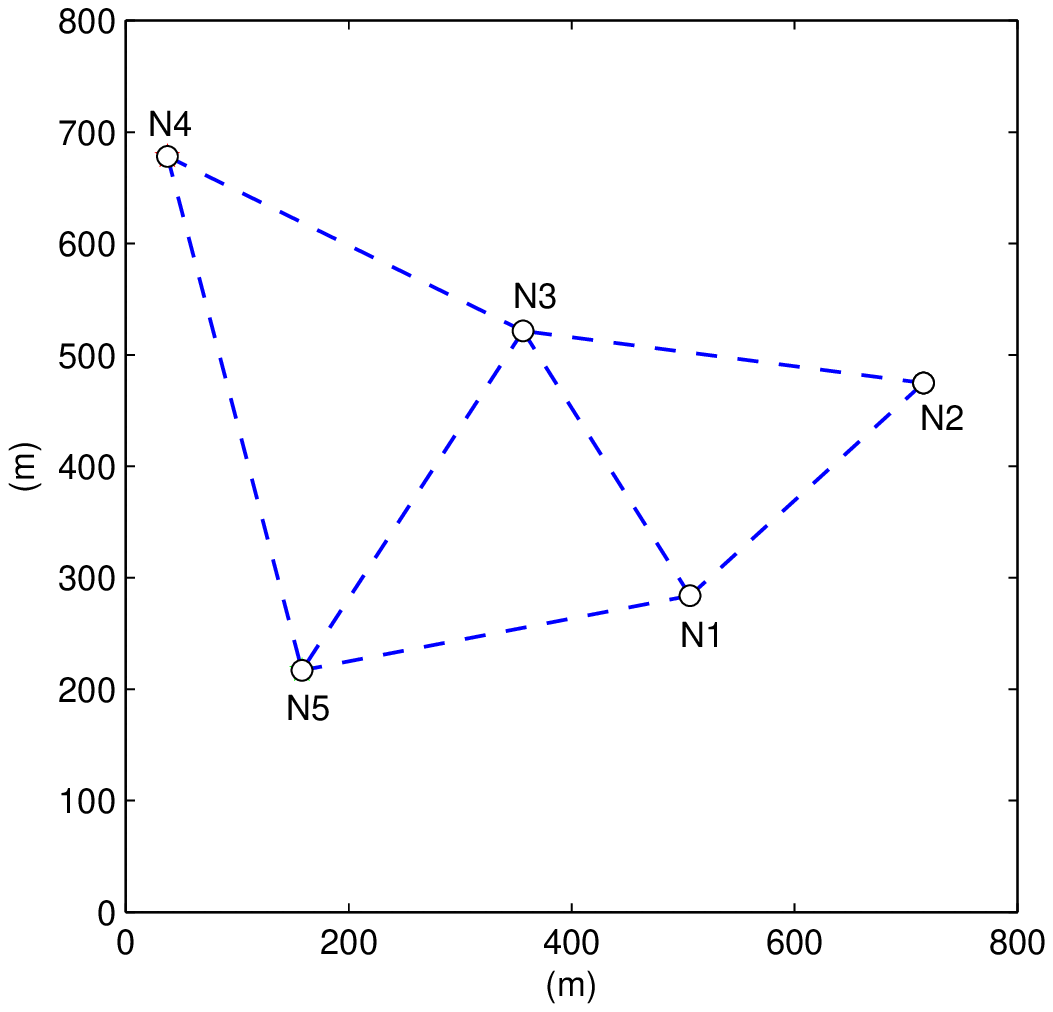}
        \caption{A five-node two-session network.} \label{fig_NetworkExample28}
    \end{minipage}%
    \hspace{.1in}
    \begin{minipage}[t]{0.32\linewidth}
        \centering
        \includegraphics[width=2.2in]{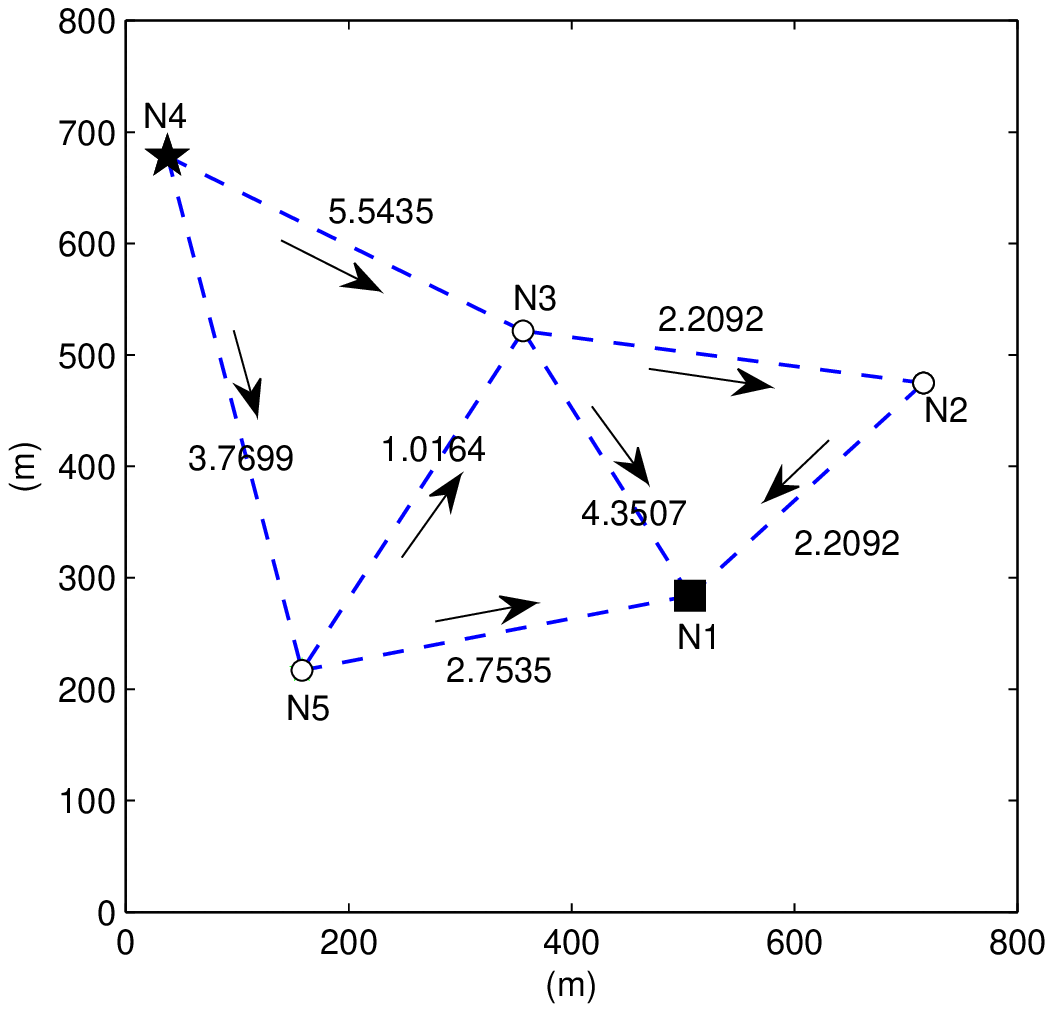}
        \caption{The optimal routing solutions for session N4 $\rightarrow$ N1 (in b/s/Hz).} \label{fig_Net28Routing_S1}
    \end{minipage}%
    \hspace{.1in}
    \begin{minipage}[t]{0.32\linewidth}
        \centering
        \includegraphics[width=2.2in]{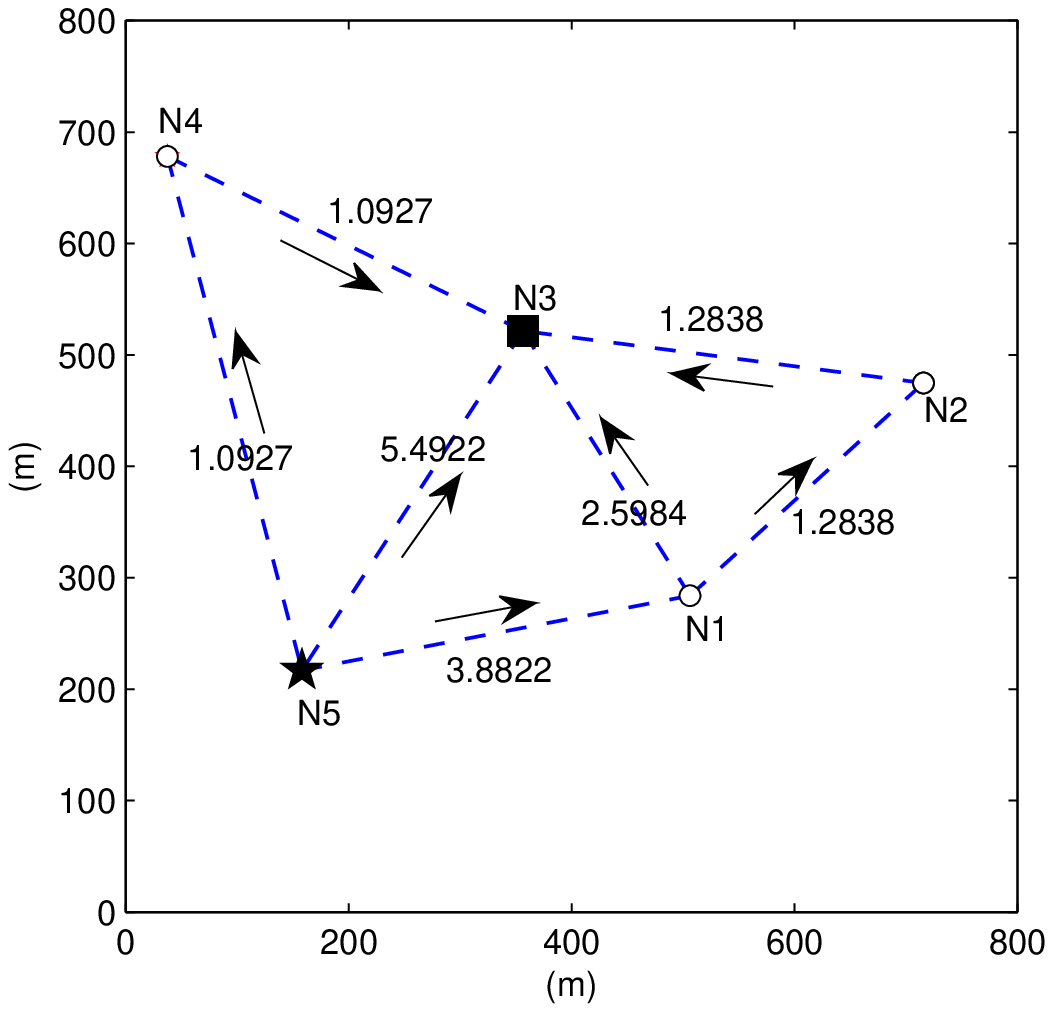}
        \caption{The optimal routing solutions for session N5 $\rightarrow$ N3 (in b/s/Hz).} \label{fig_Net28Routing_S2}
    \end{minipage}%
\vspace{-.1in}
\end{figure*}

\begin{figure}[t!]
\centering
\includegraphics[width=3.2in]{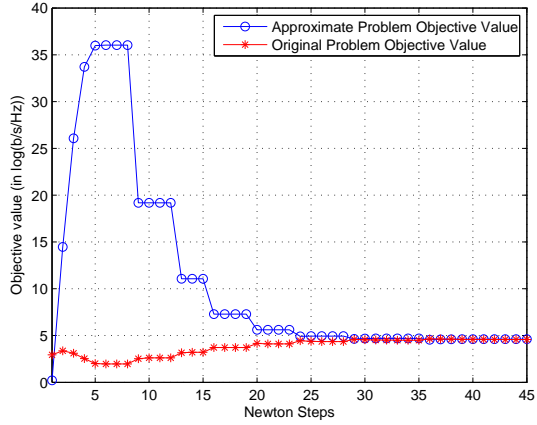}
\caption{Convergence behavior of the proposed distributed Newton algorithm for the five-node network example.} \label{fig_ConvPlot28}
\end{figure}

To show the details in our proposed distributed Newton method, we first study a five-node multi-hop wireless network as shown in Fig.~\ref{fig_NetworkExample28}.
In this network, five nodes are distributed in a square region of $800m \times 800m$.
The maximum power for each node is 100 mW.
The path-loss index is set to 3.5.
There are two sessions in the network: N4 to N1 and N5 to N3.
We adopt $\log(s_{f})$ as our utility function, which represents the so-called proportional fairness \cite{Kelly98:NUM}.
The optimal routing paths for session N4 $\rightarrow$ N1 and N5 $\rightarrow$ N3 are plotted in Fig.~\ref{fig_Net28Routing_S1} and Fig.~\ref{fig_Net28Routing_S2}, respectively.
In Fig.~\ref{fig_Net28Routing_S1} and Fig.~\ref{fig_Net28Routing_S2}, the $\bigstar$ marker and the $\blacksquare$ marker denote the source node and the destination node of each session, respectively.
The optimal session rates for N4 $\rightarrow$ N1 and N5 $\rightarrow$ N3 are 9.31334 and 10.4670 b/s/Hz, respectively.
The convergence behavior of the distributed Newton method is illustrated in Fig.~\ref{fig_ConvPlot28}, which shows the objective values of the approximating and the original problems.
It can be seen that our proposed algorithm only takes 45 Newton steps to converge, which is very efficient.

%
%

\begin{figure}[t!]
\centering
\includegraphics[width=3.2in]{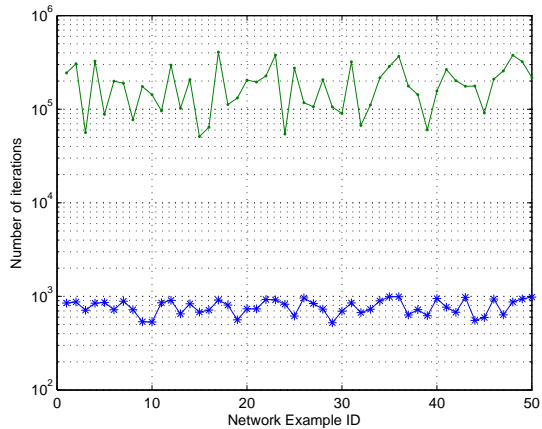}
\caption{Convergence speed comparison between our proposed algorithm and the subgradient algorithm over 50 randomly generated network examples.} \label{fig_NewtSubgComp}
\end{figure}

To further illustrate the advantage of our proposed algorithm over first-order approaches,
we randomly generate 50 network examples with 30 nodes and six sessions.
We compare the number of iterations for our proposed algorithm and the subgradient algorithm, and the results are shown in Fig.~\ref{fig_NewtSubgComp}.
For these 50 examples, the mean numbers of iterations for our distributed Newton method and the subgradient method are 779.3 and 61115.26.

\section{Conclusion} \label{sec:conclusion}

\appendix
\section{Proof of Theorem~\ref{thm:Xl_closedform}} \label{appdx:Xl_closedform}
First, note from (\ref{eqn_Xl_struc}) that $\X_{l}$ can be decomposed into a diagonal matrix with a rank-one update as follows:
\begin{equation*}
\X_{l} = \D + \frac{1}{\delta_{l}^{2}} \1 \cdot \1^{T},
\end{equation*}
where $\D$ is defined as $\D \triangleq \diag{(x_{l}^{(1)})^{-2}, \ldots, (x_{l}^{(F)})^{-2}}$.
Now, let $\u = \frac{1}{\delta_{l}^{2}} \1$ and $\v = \1$.
Then, using the Sherman--Morrison--Woodbury formula \cite{Horn_Johnson90:Mtrx_Thry}, we have that
\begin{equation} \label{eqn_Sherman_Morrison}
\X_{l}^{-1} = \D^{-1} - \frac{ \D^{-1} \u \v^{T} \D^{-1} }{1 + \v^{T} \D^{-1} \u }.
\end{equation}
Since $\D$ is diagonal, we have $\D^{-1} = \diag{(x_{l}^{(1)})^2, \ldots, (x_{l}^{(F)})^{2}}$.
The denominator of the second term in (\ref{eqn_Sherman_Morrison}) can thus be computed as follows:
\begin{align} \label{eqn_SM_denominator}
& 1 + \v^{T} \D^{-1} \u \nonumber\\
& = 1 + \frac{1}{\delta_{l}^{2}} \1^{T} \diag{(x_{l}^{(1)})^2, \ldots, (x_{l}^{(F)})^{2}} \1 \nonumber\\
&= 1 + \frac{\sum_{f=1}^{F}(x_{l}^{(f)})^{2}}{\delta_{l}^{2}}.
\end{align}
The numerator of the second term in (\ref{eqn_Sherman_Morrison}) can be computed as
\begin{align} \label{eqn_SM_nominator}
& \D^{-1} \u \v^{T} \D^{-1} = \frac{\Q}{\delta_{l}^{2}},
\end{align}
where the entries of $\Q$ are
\begin{equation*}
(\Q)_{f_{1}f_{2}} =
\begin{cases}
\big(x_{l}^{(f_{1})}\big)^{4}, & \text{if $1 \leq f_{1}=f_{2} \leq F$}, \\
\big(x_{l}^{(f_{1})} x_{l}^{(f_{2})}\big)^{2}, & \text{if $1 \leq f_{1},f_{2} \leq F, f_{1}\ne f_{2}$}.
\end{cases}
\end{equation*}
From (\ref{eqn_SM_denominator}) and (\ref{eqn_SM_nominator}), we obtain that
\begin{equation*}
\frac{ \D^{-1} \u \v^{T} \D^{-1} }{1 + \v^{T} \D^{-1} \u } = \frac{\Q}{\delta_{l}^{2} + \sum_{f=1}^{F}(x_{l}^{(f)})^{2}}.
\end{equation*}
Hence, the diagonal entries of $\X_{l}^{-1}$ can be computed as
\begin{align} \label{eqn_Xl_diag}
(\X_{l}^{-1})_{ff} &= (x_{l}^{(f)})^{2} - \frac{\big(x_{l}^{(f_{1})}\big)^{4}}{\delta_{l}^{2}+\sum_{f=1}^{F}(x_{l}^{(f)})^{2}}, \nonumber\\
 &= \big(x_{l}^{(f)}\big)^{2} \left( 1 - \frac{\big(x_{l}^{(f)}\big)^{2}}{\delta_{l}^{2}+\sum_{f=1}^{F}(x_{l}^{(f)})^{2}} \right).
\end{align}
Now, define a vector $\xh_{l} = \big[ x_{l}^{(1)}, \ldots, x_{l}^{(F)}, \delta_{l} \big]^{T}$.
Then, the right-hand-side of (\ref{eqn_Xl_diag}) can be further written as
\begin{equation} \label{eqn_Xl_diag1}
(\X_{l}^{-1})_{ff} = \big(x_{l}^{(f)}\big)^{2} \left( 1 - \frac{\big(x_{l}^{(f)}\big)^{2}}{\|\xh_{l}\|^{2}} \right).
\end{equation}
On the other hand, the off-diagonal entries of $\X_{l}^{-1}$ can be computed as
\begin{equation} \label{eqn_Xl_offdiag}
(\X_{l}^{-1})_{f_{1}f_{2}} = 0 - \frac{(x_{l}^{(f_{1})}x_{l}^{(f_{2})})^{2}}{\delta_{l}^{2} + \sum_{f=1}^{F}(x_{l}^{(F)})^{2}} = -\frac{(x_{l}^{(f_{1})}x_{l}^{(f_{2})})^{2}}{\|\xh_{l}\|^{2}}.
\end{equation}
Combining (\ref{eqn_Xl_diag1}) and (\ref{eqn_Xl_offdiag}), we obtain (\ref{eqn_Xl_closedform}).
This completes the proof.

\section{Proof of Theorem~\ref{thm:primal_newton_direct}} \label{appdx:Primal_Newton_Dir}
First, note that
\begin{equation*}
\Mt^{T} \wt^{k} =
\mtrx{{ccc}
(\bt^{(1)})^{T} & & \\
 & \ddots & \\
 & & (\bt^{(F)})^{T} \\
\hline -(\a_{1}^{(1)})^{T} & & \\
 & \ddots & \\
 & & -(\a_{1}^{(F)})^{T} \\
\hline & \vdots & \\
\hline -(\a_{L}^{(1)})^{T} & & \\
 & \ddots & \\
 & & -(\a_{L}^{(F)})^{T} \\
}
\mtrx{{c} \wt^{(1)} \\ \vdots \\ \wt^{(F)}}
=
\mtrx{{c}
(\bt^{(1)})^{T} \wt^{(1)} \\
\vdots \\
(\bt^{(F)})^{T} \wt^{(F)} \\
\hline -(\a_{1}^{(1)})^{T} \wt^{(1)} \\
\vdots \\
-(\a_{1}^{(F)})^{T} \wt^{(F)}\\
\hline \vdots \\
\hline -(\a_{L}^{(1)})^{T} \wt^{(1)} \\
\vdots \\
-(\a_{L}^{(F)})^{T} \wt^{(F)} \\
}
=
\mtrx{{c}
\widetilde{w}_{\src{1}}^{(1)} \\
\vdots \\
\widetilde{w}_{\src{F}}^{(F)} \\
\hline \widetilde{w}_{\rx{1}}^{(1)} - \widetilde{w}_{\tx{1}}^{(1)} \\
\vdots \\
\widetilde{w}_{\rx{1}}^{(F)} - \widetilde{w}_{\tx{1}}^{(F)} \\
\hline \vdots \\
\hline \widetilde{w}_{\rx{L}}^{(1)} - \widetilde{w}_{\tx{L}}^{(1)} \\
\vdots \\
\widetilde{w}_{\rx{L}}^{(F)} - \widetilde{w}_{\tx{L}}^{(F)} \\
},
\end{equation*}
where the last equality holds due to the special structure of $\bt^{(f)}$ and $\a_{l}^{(F)}$.
More specifically, notice that $\bt^{(f)}$ is simply a unit vector where all entries are zeros except for a ``1'' at the entry corresponding to the node $\src{f}$.
Thus, we have $(\bt^{(f)})^{T} \wt^{(f)} = \widetilde{w}_{\src{f}}^{(f)}$.
Likewise, $\a_{l}^{(f)}$ has two non-zero entries (a ``$1$'' corresponding to node $\tx{l}$ and a ``$-1$'' corresponding to node $\rx{l}$) or only one non-zero entry when one end point of link $l$ happens to be $\dst{f}$.
Thus, we have $-(\a_{l}^{(f)})^{T} \wt^{(f)} = \widetilde{w}_{\rx{l}}^{(f)} - \widetilde{w}_{\tx{l}}^{(f)}$ (recall that we have defined $\widetilde{w}_{\dst{f}}^{(f)}=0$).
Hence,
\begin{align*}
(\nabla f(\y^{k}) + \Mt^{T} \wt^{(k)})_i =
\begin{cases}
-tU'_{f}(s_{f}) - \frac{1}{s_{f}} + \widetilde{w}_{\src{f}}^{(f)} & \text{if $1 \leq i=f \leq F$}, \\
\frac{1}{\delta_{l}} - \frac{1}{x_{l}^{(f)}} + \widetilde{w}_{\rx{l}}^{(f)} - \widetilde{w}_{\tx{l}}^{(f)} & \text{if $i=(l+1)F + f$}.
\end{cases}
\end{align*}

Recall that $\tHc_{k}^{-1} = \diag{\S^{-1}, \X_{1}^{-1}, \ldots, \X_{L}^{-1}}$.
Thus, we have that $\Delta \y^{k} = -\tHc_{k}^{-1}(\nabla f(\y^{k}) + \Mt^{T} \wt^{(k)})$ can be partitioned into $L+1$ $F$-dimension vectors.
For the entries in the first vector, since $\S^{-1}$ is diagonal, we have
\begin{equation} \label{eqn_primal_dir1}
(\Delta \y^{k})_{i} = \frac{s_{f}(ts_{f}U'_{f}(s_{f})+1-s_{f}w_{\src{f}}^{(f)})}{1-ts_{f}^{2}U''_{f}(s_{f})}, \quad \text{if $1 \leq i=f \leq F$}.
\end{equation}
The remaining of the $F$-dimensional vectors are of the form
\begin{equation*}
-\X_{l}^{-1} \mtrx{{c}
\frac{1}{\delta_{l}} - \frac{1}{x_{l}^{(1)}} + \widetilde{w}_{\rx{l}}^{(1)} - \widetilde{w}_{\tx{l}}^{(1)} \\
\vdots \\
\frac{1}{\delta_{l}} - \frac{1}{x_{l}^{(F)}} + \widetilde{w}_{\rx{l}}^{(F)} - \widetilde{w}_{\tx{l}}^{(F)}
}, \quad l=1,\ldots,L.
\end{equation*}
Using the structural result of $\X_{l}^{-1}$ in (\ref{eqn_Xl_struc}), we have that the $i$-th entry of $\Delta \y^{k}$, where $i=(l+1)F+f$, can be computed as
\begin{align} \label{eqn_primal_dir2}
(\Delta \y^{k})_{i} &= \big( x_{l}^{(f)} \big)^{2}
\left[ \left( 1 - \frac{(x_{l}^{(f)})^{2}}{\|\xh_{l}\|^{2}} \right)
\left( \frac{1}{x_{l}^{(f)}} - \frac{1}{\delta_{l}} + \widetilde{w}_{\tx{l}}^{(f)} - \widetilde{w}_{\rx{l}}^{(f)} \right) +
\right. \nonumber\\
& \hspace{.8in} \left. \sum_{f'=1,f'\ne f}^{F} \frac{(x_{l}^{(f')})^{2}}{\|\xh_{l}\|^{2}} \left( \frac{1}{x_{l}^{(f')}} - \frac{1}{\delta_{l}} + \widetilde{w}_{\tx{l}}^{(f')} - \widetilde{w}_{\rx{l}}^{(f')} \right) \right].
\end{align}
Note that (\ref{eqn_primal_dir1}) and (\ref{eqn_primal_dir2}) are the same as (\ref{eqn_pndir1}) and (\ref{eqn_pndir2}), respectively.
The proof is complete.

\section{Proof of Lemma~\ref{lem:D_f}} \label{appdx:D_f}
First, consider the diagonal entries in $\D_{f}$.
Note that $\D_{f} = \frac{s_{f}^{2}}{t}\bt^{(f)}(\bt^{(f)})^{T} + \sum_{l=1}^{L} (x_{l}^{(f)})^{2} \a_{l}^{(f)} (\a_{l}^{(f)})^{T}$.
From Lemma~\ref{lem:alfalft}, the $i$-th diagonal entry in $\a_{l}^{(f)} (\a_{l}^{(f)})^{T}$ is equal to 1 if the corresponding node of the $i$-th entry, say $n$, is either $\tx{l}$ or $\rx{l}$.
Thus, when summing over all $l$, the number of ones is precisely given by the number of links that have node $n$ either as its transmitting node or receiving node, i.e., the links that are in either $\Out{n}$ and $\In{n}$.
Thus, we have $(\sum_{l=1}^{L} (x_{l}^{(f)})^{2} \a_{l}^{(f)} (\a_{l}^{(f)})^{T})_{ii} = \sum_{l \in \In{n}\cup\Out{n}} (x_{l}^{(f)})^{2}$.
Also, from Lemma~\ref{lem:btfbtft}, we have that the $i$-th diagonal entry is equal to 1 if $n=\src{f}$.
Hence, we have
\begin{equation*}
(\D_{f})_{ii} =
\begin{cases}
\sum_{l\in \Out{n} \cup \In{n}} (x_{l}^{(f)})^{2} + \frac{s_{f}^{2}}{t} & \text{if row $i$ corresponds to node $n$ and $n=\src{f}$}, \\
\sum_{l\in \Out{n} \cup \In{n}} (x_{l}^{(f)})^{2} & \text{otherwise},
\end{cases}
\end{equation*}
which is the same expression as in Lemma~\ref{lem:D_f}.

Next, consider the off-diagonal entries in $\D_{f}$.
Again, from Lemma~\ref{lem:alfalft}, we know that the $(i,j)$-th entry in $\a_{l}^{(f)} (\a_{l}^{(f)})^{T}$ is equal to $-1$ if the corresponding nodes of the $(i,j)$-th entry, say $n_{1}$ and $n_{2}$, are $\tx{l}$ and $\rx{l}$, or vice versa.
Thus, when summing over all $l$, the number of $-1$ entries is precisely given by the number of links that have nodes $n_{1}$ and $n_{2}$ either as their transmitting node and receiving node, i.e., the links that are in $\Gamma(n_{1},n_{2})$.
Hence, we have
\begin{equation*}
(\D_{f})_{ij} =
\begin{cases}
-\sum_{l \in \Gamma(n_{1},n_{2})}(x_{l}^{(f)})^{2} & \text{if row $i$ and column $j$ correspond to two connected nodes $n_{1}$ and $n_{2}$}, \\
0 & \text{otherwise},
\end{cases}
\end{equation*}
which is the same expression as in Lemma~\ref{lem:D_f}, and the proof is complete.

\section{Proof of Theorem~\ref{thm_Distr_Dual_Update}} \label{appdx:Distr_Dual_Update}
First, note that $\Mt \tHc_{k}^{-1} \Mt^{T} \succ 0$ because $f(\y)$ is convex.
Hence, $(\mLambda_{k}+\alpha\mOmegab) - (\alpha\mOmegab - \mOmega_{k})$ is positive definite.
Next, we check the positive definiteness of $(\mLambda_{k}+\alpha\mOmegab) + (\alpha\mOmegab - \mOmega_{k})$.
Note that
\begin{equation}
(\mLambda_{k}+\alpha\mOmegab) + (\alpha\mOmegab - \mOmega_{k}) = \mLambda_{k} + 2\alpha \mOmegab_{k} - \mOmega_{k}.
\end{equation}
From the definition of $\mLambda_{k}$, Lemma~\ref{lem:D_f}, and Lemma~\ref{lem:Dh_f}, we have that all diagonal entries in $\mLambda_{k}$ are positive.
Hence, $\mLambda_{k} \succ 0$.
On the other hand, by the definitions of $\mOmegab_{k}$ and $\mOmega_{k}$, we have that the entries of each row in $2\alpha\mOmegab_{k} - \mOmega_{k}$ satisfy
\begin{align*}
& (2\alpha \mOmegab_{k} - \mOmega_{k})_{ii} - \sum_{j\ne i} |(2\alpha \mOmegab_{k} - \mOmega_{k})_{ij}| \\
& = (2\alpha-1) \sum_{j \ne i} |(\mOmega_{k})_{ij}| > 0, \quad \text{for $\alpha > \frac{1}{2}$}.
\end{align*}
Also, it is clear from the definitions of $\mOmegab_{k}$ and $\mOmega_{k}$ that $(2\alpha \mOmegab_{k} - \mOmega_{k})_{ii} > 0$.
Thus, $2\alpha \mOmegab_{k} - \mOmega_{k}$ is diagonally dominant and hence positive definite.
Therefore, $\mLambda_{k} + 2\alpha \mOmegab_{k} - \mOmega_{k}$ is also positive definite, and the proof is complete.

\section{Proof of Proposition~\ref{prop:Alpha_Choice}} \label{appdx:Alpha_Choice}
To establish Proposition~\ref{prop:Alpha_Choice}, we need the following result \cite[Theorem 2.3]{Woznicki01:Mtrx_Split}:
\begin{lem} \label{lem:Comparison_Thm}
Let $\A = \M_{1} - \N_{1} = \M_{2} - \N_{2}$ be two splittings of $\A$, where $\A^{-1} \succeq 0$, $\M_{1}^{-1} \succeq 0$, and $\M_{2}^{-2} \succeq 0$.
If $\M_{1}^{-1} \succeq \M_{2}^{-1}$, then $\rho(\M_{1}^{-1}\N_{1}) \leq \rho(\M_{2}^{-1}\N_{2})$.
\end{lem}
Now, for $\frac{1}{2} < \alpha_{1} \leq \alpha_{2}$, since $\mLambda_{k} + \alpha\mOmegab_{k}$ is diagonal, we have that
\begin{align} \label{eqn_Alpha_Matrix_Subtract}
& (\mLambda_{k} + \alpha_{2}\mOmegab_{k})_{ii} - (\mLambda_{k} + \alpha_{1}\mOmegab_{k})_{ii} \nonumber\\
&= (\alpha_{2}-\alpha_{1}) \sum_{j\ne i} |(\mOmega_{k})_{ij}| > 0.
\end{align}
Also, since $((\mLambda_{k} + \alpha\mOmegab_{k})^{-1})_{ii} = 1/(\mLambda_{k} + \alpha\mOmegab_{k})_{ii}$ (from the diagonal property again), Eq. (\ref{eqn_Alpha_Matrix_Subtract}) implies that $(\mLambda_{k} + \alpha_{1}\mOmegab_{k})^{-1} \succeq (\mLambda_{k} + \alpha_{2}\mOmegab_{k})^{-1}$.
Thus, Proposition~\ref{prop:Alpha_Choice} simply follows from Lemma~\ref{lem:Comparison_Thm}, and the proof is complete.

\section{Proof of Theorem~\ref{thm:Distr_Dual_Update_Exp}} \label{appdx:Distr_Dual_Update_Exp}
To show (\ref{eqn_Distr_Dual_Update_Exp}), we need to compute the element-wise expansion of (\ref{eqn_Distr_Dual_Update}).
First, note that $(\mLambda_{k}+\alpha\mOmegab_{k})$ is diagonal, and so its inverse can be easily computed by taking the inverse of each diagonal entry.
Therefore, we start by computing each diagonal entry in $(\mLambda_{k}+\alpha\mOmegab_{k})$.
To this end, we first define an index function $\beta_{f}(n)$, $n \ne \dst{f}$, as follows:
\begin{equation}
\beta_{f}(n) \triangleq
\begin{cases}
n & \text{if $n<\dst{f}$}, \\
n-1 & \text{if $n>\dst{f}$}.
\end{cases}
\end{equation}
Since $\mLambda_{k}$ contains the main diagonal of $\Mt \tHc_{k}^{-1} \Mt^{T}$, from Theorem~\ref{thm:MHMT}, we obtain that
\begin{equation} \label{eqn_mLambda_ii}
(\mLambda_{k})_{ii} =
\begin{cases}
\sum_{\Phi(n)} (x_{l}^{(f)})^{2} \Big( 1- \frac{(x_{l}^{(f)})^{2}}{\|\xh_{l}\|^{2}} \Big) + \frac{1}{-tU''_{f}(s_{f})+\frac{1}{(s_{f})^{2}}} & \text{if $n = \src{f}$}, \\
\sum_{\Phi(n)} (x_{l}^{(f)})^{2} \Big( 1- \frac{(x_{l}^{(f)})^{2}}{\|\xh_{l}\|^{2}} \Big) & \text{if $n \ne \src{f}$},
\end{cases}
\end{equation}
where the index $i$ satisfies $i=(f-1)(N-1)+\beta_{f}(n)$.
Notice that each diagonal entry in $\mOmegab_{k}$ is the row sum of non-diagonal entries in $\Mt \tHc_{k}^{-1} \Mt^{T}$.
Thus, from Theorem~\ref{thm:MHMT}, we obtain that
\begin{equation} \label{eqn_mOmegab_ii}
(\mOmegab)_{ii} = \sum_{l\in \Phi(n)\backslash \Psi(n,f)} (x_{l}^{(f)})^{2} \Big( 1- \frac{(x_{l}^{(f)})^{2}}{\|\xh_{l}\|^{2}} \Big) + \sum_{f'=1,\ne f}^{F} \sum_{l\in \Psi(n,f')} \frac{(x_{l}^{(f)}x_{l}^{(f')})^{2}}{\|\xh_{l}\|^{2}}.
\end{equation}
Therefore, using the indicator function $\ind_{\Psi(n,f)}$ and adding (\ref{eqn_mLambda_ii}) and (\ref{eqn_mOmegab_ii}), we obtain that
\begin{equation*}
(\mLambda_{k}+\alpha\mOmegab_{k})_{ii} =
\begin{cases}
\sum_{l\in \Phi(n)}[1+\alpha(1-\ind_{\Psi(n,f)}(l))] (x_{l}^{(f)})^{2}\Big(1-\frac{(x_{l}^{(f)})^{2}}{\|\xh_{l}\|^{2}} \Big) + & \\
\hspace{.1in} \sum_{f'=1,\ne f}^{F}\Big( \sum_{l \in \Psi(n,f')} \frac{\alpha(x_{l}^{(f)}x_{l}^{(f')})^{2}}{\|\xh_{l}\|^{2}} \Big) & \text{if $n\ne \src{f}$}, \\
\sum_{l\in \Phi(n)}[1+\alpha(1-\ind_{\Psi(n,f)}(l))] (x_{l}^{(f)})^{2}\Big(1-\frac{(x_{l}^{(f)})^{2}}{\|\xh_{l}\|^{2}} \Big) + & \\
\hspace{.1in} \sum_{f'=1,\ne f}^{F}\Big( \sum_{l \in \Psi(n,f')} \frac{\alpha(x_{l}^{(f)}x_{l}^{(f')})^{2}}{\|\xh_{l}\|^{2}} \Big) + \frac{1}{-tU''_{f}(s_{f})+\frac{1}{(s_{f})^{2}}}& \text{if $n=\src{f}$},
\end{cases}
\end{equation*}
which is the same as the definition of $U_{n}^{(f)}(k)$ in (\ref{eqn_Unfk}).

Next, consider the entries in $(\alpha\mOmegab_{k} - \mOmega_{k})\wt^{k}$.
Recall from Theorem~\ref{thm:MHMT} that the matrix $\Mt\tHc_{k}^{-1}\Mt^{T}$ has a partitioned matrix structure.
Thus, the vector $(\alpha\mOmegab_{k}-\mOmega_{k})\wt^{k}$ can be partitioned into $F$ blocks, where each block is of the form
\begin{equation}
((\alpha\mOmegab_{k} - \mOmega_{k})\wt^{k})_{f} = -\R_{f}\wt^{f} + \sum_{f'=1,\ne f}^{F} \G_{ff'}\wt^{(f')}, \quad f=1,\ldots,F,
\end{equation}
where $\R_{f}$ is obtained by replacing the main diagonal of $\D_{f}-\Dh_{f}$ with the corresponding entries in $-\alpha\mOmegab_{k}$.
Hence, by computing the entries in $-\R_{f}\wt^{f}$ and noting the special structure in $\R_{f}$, where it only contains entries $1$, $-1$, and $0$, we have
\begin{align*}
(-\R_{f}\wt^{f})_{n} &=
\sum_{l\in \In{n}} (x_{l}^{(f)})^{2} \Big(1-\frac{((x_{l}^{(f)})^{2})}{\|\xh_{l}\|^{2}}\Big)(\wts_{\tx{l}}^{(f)} -\alpha\wts_{\rx{l}}^{(f)}) + \nonumber\\
& \hspace{.2in} \sum_{l\in \Out{n}\backslash \Psi(n,f)} (x_{l}^{(f)})^{2} \Big(1-\frac{((x_{l}^{(f)})^{2})}{\|\xh_{l}\|^{2}}\Big)(\wts_{\rx{l}}^{(f)} - \alpha\wts_{\tx{l}}^{(f)}) - \nonumber\\
& \hspace{.2in} \sum_{f'=1,\ne f}^{F} \Big( \sum_{l \in \Psi(n,f')} \frac{\alpha(x_{l}^{(f)}x_{l}^{(f')})^{2}}{\| \xh_{l} \|^{2}} \Big)\wts_{n}^{f},
\end{align*}
which is the same as the definition of $V_{n,1}^{(f)}(k)$ in (\ref{eqn_Vn1fk}).
Similarly, by computing the entries in $\sum_{f'=1,\ne f}^{F} \G_{ff'}\wt^{(f')}$, we have
\begin{equation*}
\Big( \sum_{f'=1,\ne f}^{F} \G_{ff'}\wt^{(f')} \Big)_{n} =
\sum_{f'=1,\ne f}^{F} \Big(
\Big(\sum_{l \in \Out{n}} \frac{(x_{l}^{(f)}x_{l}^{(f')})^{2}}{\|\xh_{l}\|^{2}} -
\sum_{l \in \In{n}} \frac{(x_{l}^{(f)}x_{l}^{(f')})^{2}}{\|\xh_{l}\|^{2}} \Big)
(\wts_{\tx{l}}^{(f')} - \wts_{\rx{l}}^{(f')})
\Big),
\end{equation*}
which is the same as the definition of $V_{n,2}^{(f)}(k)$ in (\ref{eqn_Vn2fk}).

Finally, consider the term $\Mt \tHc_{k}^{-1} \nabla f(\yt^{k})$.
Note that $\Mt \tHc_{k}^{-1} \nabla f(\yt^{k})$ can be decomposed into
\begin{equation*}
\Mt \tHc_{k}^{-1} \nabla f(\yt^{k}) = \Bt \S^{-1} \nabla_{\s} f(\yt^{k}) + \sum_{l=1}^{L} -\A_{l} \X_{l}^{-1} \nabla_{\x_{l}} f(\yt^{k}),
\end{equation*}
where $\s \triangleq [s_{1},\ldots,s_{F}]^{T}$ and $\x_{l} \triangleq [x_{l}^{(1)},\ldots,x_{l}^{(F)}]^{T}$.
Accordingly, consider first the term $\Bt \S^{-1} \nabla_{\s} f(\yt^{k})$.
Using the diagonal structure of $\Bt$ and $\S$, it is easy to obtain that
\begin{equation*}
(\Bt \S^{-1} \nabla_{\s} f(\yt^{k}))_{n}^{(f)} =
\begin{cases}
\frac{s_{f}(1+ts_{f}U'_{f}(s_{f}))}{ts_{f}^{2}U''_{f}(s_{f})-1} & \text{if $n=\src{f}$},\\
0 & \text{otherwise}.
\end{cases}
\end{equation*}
Recalling that $\tHc_{k}^{-1}$ can be decomposed into a diagonal matrix and a rank-one update matrix, we have
\begin{align*}
-\A_{l}\X_{l}^{-1} \nabla_{\x_{l}} f(\yt^{k}) &= -\A_{l} \diag{(x_{l}^{(1)})^{2},\ldots,(x_{l}^{(F)})^{2}}
\mtrx{{c}
\frac{1}{\delta_{l}}-\frac{1}{x_{l}^{(1)}} \\
\vdots \\
\frac{1}{\delta_{l}}-\frac{1}{x_{l}^{(F)}}
} + \\
& \hspace{.2in} \frac{1}{\|\xh_{l}\|^{2}}\A_{l}
\mtrx{{ccc}
(x_{l}^{(1)})^{4} & \cdots & (x_{l}^{(1)}x_{l}^{(F)})^{2} \\
\vdots & \ddots & \vdots \\
(x_{l}^{(F)}x_{l}^{(1)})^{2} & \cdots & (x_{l}^{(F)})^{4} \\
}
\mtrx{{c}
\frac{1}{\delta_{l}}-\frac{1}{x_{l}^{(1)}} \\
\vdots \\
\frac{1}{\delta_{l}}-\frac{1}{x_{l}^{(F)}}
}.
\end{align*}
Hence, computing each term in the above decomposition, then adding $\Bt \S^{-1} \nabla_{\s} f(\yt^{k})$, and then summing over all $l$, we obtain that
\begin{equation*}
(\Mt \tHc_{k}^{-1} \nabla f(\yt^{k}))_{n}^{(f)} =
\begin{cases}
\Big(1-\frac{x_{l}^{(f)}}{\delta_{l}}\Big)
\Big[
\sum_{l\in \Out{n}} \Big( 1-\sum_{f'=1}^{F} \frac{(x_{l}^{(f)})^{2}}{\|\xh_{l}\|^{2}} x_{l}^{(f')} \Big) - & \\
\hspace{.3in} \sum_{l\in \In{n}} \Big( 1-\sum_{f'=1}^{F} \frac{(x_{l}^{(f)})^{2}}{\|\xh_{l}\|^{2}} x_{l}^{(f')} \Big)
\Big] & \text{if $n \ne \src{f}$}, \\
\sum_{l\in \Out{n}} \Big( 1-\sum_{f'=1}^{F} \frac{(x_{l}^{(f)})^{2}}{\|\xh_{l}\|^{2}} x_{l}^{(f')} \Big) - & \\
\hspace{.3in} \sum_{l\in \In{n}} \Big( 1-\sum_{f'=1}^{F} \frac{(x_{l}^{(f)})^{2}}{\|\xh_{l}\|^{2}} x_{l}^{(f')} \Big)
\Big] + \frac{s_{f}(1+ts_{f}U'_{f}(s_{f}))}{ts_{f}^{2}U''_{f}(s_{f})-1} & \text{if $n = \src{f}$}, \\
\end{cases}
\end{equation*}
which is the same as the definition of $W_{n}^{(f)}(k)$ as in (\ref{eqn_Wnfk}).
Thus, the result in (\ref{eqn_Distr_Dual_Update_Exp}) simply follows from Theorem~\ref{thm_Distr_Dual_Update}, and the proof is complete.

\section{Proof of Proposition~\ref{prop:Distr_Newton_Decre}} \label{appdx:Distr_Newton_Decre}
Define the following two vectors: $\Delta \s \triangleq [\Delta s_{1},\ldots, \Delta s_{F}]^{T}$ and $\Delta \x_{l} = [x_{l}^{(1)},\ldots,x_{l}^{(F)}]^{T}$.
Also, from the decomposable structure of $\tHc_{k}$, we have
\begin{align}
(\Delta \yt)^{T} \tHc_{k} \Delta \yt = (\Delta \s)^{T} \S \Delta \s +
\sum_{l=1}^{L} (\Delta \x_{l})^{T} \X_{l} \Delta \x_{l}.
\end{align}
Now, consider first $(\Delta \s)^{T} \S \Delta \s$, which, due to the diagonal structure of $\tHc_{k}$, can be simply computed as
\begin{equation} \label{eqn_Distr_Newton_Decre_1}
(\Delta \s)^{T} \S \Delta \s = \sum_{f=1}^{F} (\Delta s_{f})^{2} \Big(-tU''_{f}(s_{f})+\frac{1}{(s_{f})^{2}}\Big).
\end{equation}
Next, we consider $\sum_{l=1}^{L} (\Delta \x_{l})^{T} \X_{l} \Delta \x_{l}$.
Recall that $\X_{l}$ can be further decomposed into a diagonal matrix plus a rank-one update matrix.
Thus, we have
\begin{align} \label{eqn_Distr_Newton_Decre_2}
& \sum_{l=1}^{L} (\Delta \x_{l})^{T} \X_{l} \Delta \x_{l} \nonumber\\
& = \sum_{l=1}^{L}
\mtrx{{ccc}x_{l}^{(1)} & \cdots & x_{l}^{(F)}}
\left(
\mtrx{{ccc}
\frac{1}{(x_{l}^{(1)})^{2}} & & \\
 & \ddots & \\
 & & \frac{1}{(x_{l}^{(F)})^{2}}
} +
\frac{1}{\delta_{l}} \1\cdot\1^{T}
\right)
\mtrx{{c}x_{l}^{(1)} \\ \vdots \\ x_{l}^{(F)}} \nonumber\\
&= \sum_{l=1}^{L} \bigg[
\sum_{f=1}^{F} \Big( \frac{\Delta x_{l}^{(f)}}{x_{l}^{(f)}} \Big)^{2}
+\frac{1}{\delta_{l}} \Big( \sum_{f=1}^{F} \Delta x_{l}^{(f)} \Big)^{2}
\bigg].
\end{align}
Thus, adding (\ref{eqn_Distr_Newton_Decre_1}) and (\ref{eqn_Distr_Newton_Decre_2}) gives the desired result in Proposition~\ref{prop:Distr_Newton_Decre}, and the proof is complete.


\end{document}